\documentclass[twocolumn]{aastex6}

\usepackage{natbib}
\usepackage{rotating}
\usepackage{amsmath}
\usepackage{graphicx}
\usepackage{textcomp}
\usepackage{color}
\usepackage{tabularx}
\usepackage{appendix}

\begin{document}

\title{A close-up view of the young circumbinary disk HD~142527}
\author {Y. Boehler$^1$, E. Weaver$^1$, A. Isella$^1$, L. Ricci$^{1,2}$, C. Grady$^3$, J. Carpenter$^4$ and L. Perez$^5$}
\affil{$^1$Rice University, Department of Physics and Astronomy, Main Street, 77005 Houston, USA \\
       $^2$Harvard-Smithsonian Center for Astrophysics, 60 Garden Street, Cambridge, MA 02138, USA \\
       $^3$Exoplanets and Stellar Astrophysics Laboratory, NASA Goddard Space Flight Center, Greenbelt, MD, USA \\
       $^4$Joint ALMA Observatory (JAO), Alonso de Cordova 3107 Vitacura - Santiago de Chile, Chile \\
       $^5$Max-Planck-Institute for Astronomy, Bonn, Germany }

\begin{abstract}

We present ALMA observations of the 0.88~millimeter dust continuum, $^{13}$CO, and C$^{18}$O J=3-2 line emission of the 
circumbinary disk HD~142527 at a spatial resolution of $\sim$ 0.25\arcsec. This system is characterized by a large central cavity 
of roughly 120 AU in radius, and asymmetric dust and gas emission. By comparing the observations with theoretical models, we find 
that the azimuthal variations in gas and dust density reach a contrast of 54 for dust grains and 3.75 for CO molecules, with an extreme 
gas-to-dust ratio of 1.7 on the dust crescent. We point out that caution is required in interpreting continuum subtracted maps of 
the line emission as this process might result in removing a large fraction of the line emission. Radially, we find that both the 
gas and dust surface densities can be described by Gaussians, centered at the same disk radius, and with gas profiles wider than 
for the dust. These results strongly support a scenario in which millimeter dust grains are radially and azimuthally 
trapped toward the center of a gas pressure bump. Finally, our observations reveal a compact source of continuum and CO emission 
inside the dust depleted cavity at $\sim$ 50 AU from the primary star. The kinematics of the CO emission from this region is different 
from that expected from material in Keplerian rotation around the binary system, and might instead trace a compact disk around a 
third companion. Higher angular resolution observations are required to investigate the nature of this source.

\end{abstract}

%\keywords{Magnetohydrodynamics(MHD) -- Stars: formation-- Protoplanetary disks}

\section{Introduction}

Planets are thought to form in a few million years in dusty and gaseous disks found around pre-main sequence stars 
\citep{Hais2001}. Starting from micron size, dust grains are expected
to coagulate to millimeter and centimeter sizes by Brownian motion, electrostatic forces, and collisions at low
relative velocities (see, e.g., the review by \citealt{Blum2008}). Growth to larger sizes faces more theoretical
difficulties. First, large grains are subject to fast radial migration towards the star due to the friction with gas
rotating at a slightly sub-Keplerian velocity \citep{Weid1977,Naka1986}. Secondly, pebbles are more prone
to bounce and fragment, instead of growing by collisions \citep{Brau2008,Zsom2010}.
A solution is to invoke localized gas pressure modulations that trap solids \citep{Whip1972}, permitting them
to grow to much larger sizes \cite[see, also,][]{Pini2012}. In proximity of such ``pressure bumps",  the pressure
gradient driving the drag force is not directed towards the star, but towards the pressure bump itself, therefore
inverting the radial migration of dust grains. As a consequence, pressure bumps are expected to
trap dust particles  and perhaps catalyze the formation of planetesimals.

A particular class of disks called ``millimeter-bright transitional disks" are believed to present strong radial
pressure bumps which manifest themselves in the form of dust annuli and dust depleted cavities of tens of AU in 
radius.  According to \cite{Andr2011}, such disks represent more than 26$\%$ of the
upper quartile in millimeter luminosity of the disks. Several of these disks have now been studied and resolved 
at millimeter wavelengths with the Sub-Millimeter Array (SMA) \citep[e.g.][]{Brow2007, Isel2010, Andr2011}, the 
Combined Array for Research in Millimeter-wave Astronomy(CARMA) \citep{Isel2012, Isel2014}, the Plateau de Bure Interferometer 
(PdBI) \citep[e.g.][]{Piet2006}, and more recently
with the Atacama Large Millimeter and submillimeter Array (ALMA) \citep[e.g.][]{vdM2013, Dutr2014, Pere2014, Casa2015-2, Zhang2014}.
Observations have shown that while millimeter grains in the central cavity are depleted by a factor $>$ 100-1000,
the molecular gas is reduced by a much smaller factor \citep{vdM2015}. This result supports the idea that
millimeter grains are trapped at a large distance from the star while gas flows toward, and eventually accretes onto, the central star.
The large size of the observed structures suggests that they might form from the gravitational interaction between the
circumstellar disk and stellar or planetary mass companions \citep{Zhu2011, Dods2011, Fung2014}. This interaction is indeed
expected to create an annular pressure bump outside the orbit of the companion capable of trapping dust particles \citep{Pini2012-2}.

Observations at radio wavelengths also revealed horseshoe asymmetries in dust density in several of these
objects \citep{vdM2013,Fuka2013,Pere2014}. The current interpretation is that these structures probe dust particles
azimuthally trapped by large anticyclonic vortices  \citep{Barg1995,Zhu2014}. As for the radial trap, at the typical
densities of protoplanetary disks, millimeter grains are the particles most efficiently trapped by these structures.
Anticyclonic vortices are predicted to originate from Rossby-wave instabilities excited by sharp density gradient
at the outer edge of gas gaps opened by companions \citep{Love1999}, or changes in viscosity
\citep{Rega2012, Lyra2015}. More recently, \cite{Owen2016} also proposed that baroclinic instabilities could create long
lived vortices.

Two transitional disks were intensely studied for their remarkable properties: Oph IRS 48 for its highest
azimuthal contrast in dust flux density ($\leq$ 130) \citep{vdM2013}, and HD~142527 for having the second highest density
contrast  ($\sim  20$) and a very large radius allowing study of the disk with great detail.
In this work, we present recent ALMA observations of HD~142527 which reveal new details of the dust and gas emission and 
kinematics. HD~142527 is a 2-5 Myr binary system \citep{Fuka2006, Laco2016} at a distance of 156 pc $\pm$ 7.5 
pc\footnote{This distance is based on the stellar 
parallax measured by  the Gaia space telescope \citep[][]{Gaiaa2016,Gaiab2016}}. The primary star is an Herbig FeIIIe star 
with a mass of 2.4 $M_\odot$ \citep{Verh2011} which is accreting material at the rate of about $10^{-7}$ M$_\odot$ 
yr$^{-1}$ \citep{Garc2006, Mend2014}. The secondary star has a mass between 0.1-0.3 $M_\odot$, an orbital radius 
between 15-20 AU, and a mass accretion rate of $\mathrm{\sim 6 \times 10^{-10} M_{\odot} \: yr^{-1}}$ 
\citep{Bill2012, Clos2014, Laco2016}. The binary system is at the center of a large elliptical dust cavity of about 120 AU in radius and 
of an asymmetric disk with 
a dense dust crescent on the north side \citep{Ohas2008,Casa2015-2}. In infrared, dust is visible up to an orbital radius of 
300 AU and traces several spiral arcs \citep{Casa2012, Cano2013, Fuka2006,Rodi2014}. In addition to the circumbinary disks, 
the primary star appears to be surrounded by a much smaller dusty disk \citep{Mari2015}. From the morphology of the 
near-infrared scattered light emission, more precisely from the detection of two azimuthal minima in the intensity 
interpreted as shadows casted by an inner disk, \cite{Mari2015} were able to deduce that the circumprimary disk has an 
inclination differing by 70 degrees with respect the circumbinary disk.

We imaged HD~142527 in the 0.88~mm dust continuum emission as well as in the $^{13}$CO and C$^{18}$O J=3-2 lines
at a spatial resolution of about 30-40 AU, about a factor of two better than previous observations \citep{Fuka2013}. 
These observations reveal the morphology of the gas and dust distribution and enable us to
constrain the temperature and kinematics of the disk. They also reveal dust and gas emission arising from a compact source
locate inside the dust depleted cavity, at a separation of about 50 AU from the primary star. The observations and the data
reduction are presented in Section \ref{sec:obs}, while the morphology of the gas and dust emission is presented in Section \ref{sec:res}.
In section \ref{sec:analysis}, we constrain the radial distribution of dust and gas by comparing the observations to theoretical disk 
models. In section \ref{sec:discu}, we discuss our results in the framework of the dust trapping model. A summary of our findings is
presented in section \ref{sec:conclu}.

%LkCa 15  pietu 2006  IRAM PdBI array
%GM Aurigae Hughes 2009 Submillimeter Array (SMA)
%MWC 758 Andrews 2011    Submillimeter Array (SMA),    MWC758      Isella et al. (2010b)
%SAO 206462 (HD 135344 B) Andrews 2011, Brown et al. 2009
%LkHα 330 Andrews 2011, Brown et  al. (2008)
%SR 21
%UX Tau A: Andrews 2011
%SR 24 S : Andrws et al 2011
%DoAr 44: Andrews et al. (2009)

%%%%%%%%%%%%%%%%%%%%%%%%%%%%%%%%%%%%%%%%%%%%%%%  

\section{Observations and data reduction}
\label{sec:obs}
We obtained ALMA cycle 1 data of the HD~142527 system as part of the project 2012.1.00725.S. The observations 
were carried on 2015 June 13th in dual polarization with a total time of $\sim$ 2.9 hours and $\sim$ 1.1 hour on target, 
using 37 antennas on projected baselines between 35 and 770 meters. 
Our data contain 4 GHz bandwidth in continuum in the frequency range $\sim$ 340-344 GHz with channel width of 15.625 MHz.
We also observed the two emission lines $^{13}$CO J = 3-2 and C$^{18}$O J = 3-2, with rest frequencies 330.58797 GHz and 
329.33055 GHz respectively, in two different spectral windows of 234 MHz bandwidth and channel width of 130 kHz after Hanning 
smoothing ($\sim$ 0.11 km s$^{-1}$). 

The bandpass was calibrated by observing J1427-4206, the flux calibration was obtained by observing Titan, 
and the phase calibration was obtained by observing J1604-4441 about every 8 minutes. The data were
calibrated using the ALMA pipeline with the version 4.3.1 of CASA. We self-calibrated the visibilities for the scientific 
source by using an interval time of 30 seconds for the phases and equal to the lengths of the scan ($\sim$ 5 minutes) for 
the amplitude. The image deconvolution was performed using multi-scale cleaning with scale factors of 0, 6, 18 times the 
pixel size (0.04 \arcsec). The self-calibration improved the peak signal-to-noise by a factor of 
$\sim$ 4 in the continuum. Finally, we applied the complex gains derived from the continuum to the emission lines 
and subtracted the continuum with the task uvcontsub. 

The final map of the continuum was generated using a Briggs factor of -2 which delivers a beam size of 0.21$''\times$0.26$''$ 
(PA = -59$^\circ$), or 32.8 AU $\times$ 40.6 AU assuming a distance of 156 pc for the system. The peak intensity is 0.132 Jy 
beam$^{-1}$, and the rms is 115 $\mu$Jy beam$^{-1}$, giving a peak signal-to-noise ratio of $\sim$ 1150. For the lines, we used 
a Briggs factor of 0.5 which gives an angular resolution of 
0.27$''\times$0.31$''$ (PA = -62$^\circ$), or 42.1 AU $\times$ 48.4 AU, and an rms noise of 5.9 and 7.5 mJy beam$^{-1}$ 
per channel for $^{13}$CO J=3-2 and C$^{18}$O J=3-2, respectively.

% the dust subtraction was largely improved. 
%The peak emission is 0.132 Jy/beam for Briggs = -2 (0.180 Jy/beam for briggs = 0.5) on the dust map and 0.321 
%$Jy/beam.km.s^{-1}$ for $^{13}$CO J=3-2 and 0.167 $Jy/beam.km.s^{-1}$ for C$^{18}$O J=3-2. Compared to the noise, this gives us 
%a signal to noise of 1200 with Briggs = -2 and (noise of 112 Jy/beam) and $\sim$ 2600 with Briggs = 0.5 (noise of 65 
%Jy/beam). For the lines, the peak signal to noise is is 40 for $^{13}$CO and 25 for C$^{18}$O. 5.9 mJy/beam per channel for the 
%$^{13}$CO and 7.8 mJy/beam/channel for C$^{18}$O J=3-2.

\section{Results}
\label{sec:res}
\subsection{Disk morphology}
\subsubsection{Dust emission}
\label{sec:Dust}

\begin{figure*}
  \includegraphics[angle=-90,width=\textwidth]{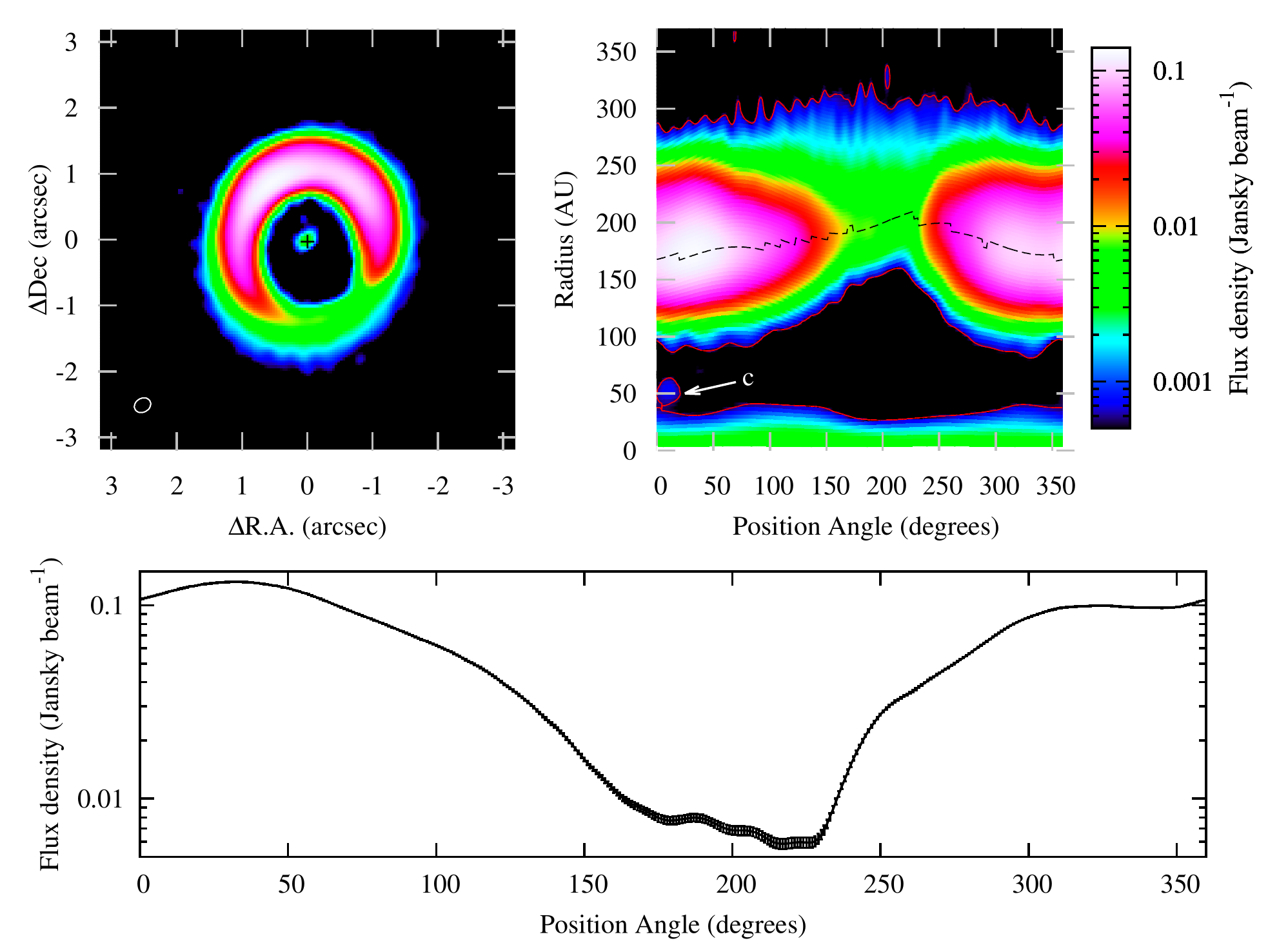}\\
  \caption{Top left: Map of the dust continuum emission recorded at a frequency of 342 GHz. The FWHM of the synthesized beam 
  is 0.21\arcsec$\times$0.26\arcsec and is indicated by the white ellipse. Top right: Dust emission as a function of 
  the orbital radius and of the position angle (PA) measured from the North to the East. The orbital radius has been calculated 
  by assuming a disk distance of 156 pc and an inclination of 27$^\circ$. The dashed line indicates the disk crest, 
  defined as the line that connects the peaks of the 
  emission measured at each position angle. The red lines indicates the 5 $\sigma$  intensity contours, corresponding 
  to 0.57 mJy beam$^{-1}$. Bottom: Peak emission on the crest as a function of the PA.}
  \label{Fig:mom0D}
\end{figure*}

The map of the 342 GHz ($\lambda$ $=$ 0.88 mm) dust thermal emission is shown on Figure~\ref{Fig:mom0D}. It unveils a 
large eccentric crescent extending radially up to $\sim$ 300 AU as previously observed by \cite{Fuka2013} and \cite{Casa2015-2}. The 
total flux density is about 3.31 Jy and is dominated by the emission coming from the northern side of the disk. 

The morphology of the dust crescent can be described by drawing the crest for the dust emission, defined as the line 
that connects the peaks of the flux density measured as a function of the position angle (black dashed line on right panel). 
The radius of the crest is calculated by assuming that the center of the disk is pinpointed by the compact source detected 
inside the dust cavity, which is coincident with the Herbig star position and with the center of rotation of the disk 
(see section \ref{sec:dyn}). The crest has a maximum of $\sim$ 132 mJy beam$^{-1}$ at a Position Angle (PA) of 30\arcdeg\ and 
a minimum of $\sim$ 6 mJy beam$^{-1}$ at PA = 220\arcdeg\, resulting in a flux density contrast of 22 between the northern and 
southern side of the disk 
and a peak-to-noise ratio in the image of 1150. The radial position of the azimuthal maximum is 165 AU while the radial position 
of the azimuthal minimum is 205 AU. Along the azimuthal direction that intersect the minimum and maximum of the crest, the 
radial profile of the continuum emission is in first approximation a Gaussian with a width of 60-80 AU. The radial profile of the 
dust emission is discussed in more details in Section~\ref{sec:detailed}. The inner and outer radius of the circumbinary dust emission 
is measured based on the $5\sigma$ intensity contours (red solid lines in the top-right panel). The inner edge varies with the 
azimuth and is located between 80 AU and 155 AU. The radius of the outer edge also depends on the azimuth and varies 
between 250 AU and 300 AU. 

In first approximation, the shape of the crest can be described by an ellipse characterized by a major axis of 185 AU, 
an eccentricity of 0.11, and a position angle of the major axis of 205$^\circ$. Using the same PA for the major axis of the ellipse, the 
best fit gives a  semi-major axis of 120 AU along with an eccentricity of 0.29 for the inner edge and a semi-major axis of 300 AU and 
an eccentricity of 0.050 for the outer edge. The eccentricity in the outer disk then decreases with the distance to the central stars. 
The periapsis and apoapsis position angle of the cavity, crest and outer edge are the same and are aligned with the dust peak emission 
on the dust crescent and with the dust minimum on the south-east side of the disk. As shown on the bottom panel of Figure~\ref{Fig:mom0D}, 
the crest also has two local minima of 97.2 mJy beam$^{-1}$ and 7.8 mJy beam$^{-1}$ at PA = 340\arcdeg\ 
and PA = 180\arcdeg\, respectively. These local minima are also visible in the CO moment 0 maps discussed in the next section 
\ref{sec:COint}.

The observations also reveal a source of compact emission inside the dust cavity with a peak flux density of 4.1 mJy beam$^{-1}$. 
Its position corresponds to the center of rotation of the disk as measured by the CO observations discussed in section \ref{sec:dyn}. 
The emission is unresolved at the angular resolution of our observations implying a radial extent less than 30 AU in diameter. A flux density 
of 80 $\mu$Jy beam$^{-1}$ was detected at 34 GHz at the same position by \cite{Casa2015-2}. At these frequencies, the signal might 
be contaminated 
by free-free emission. In order to estimate the amount of possible free-free emission in our ALMA data, we assume the emission observed at 
34 GHz is entirely free-free  and has a spectral index $\alpha$  ($I_\nu$ $\propto$ $\nu^{\alpha}$) equal to 0.4, being this value the 
steepest index predicted by theoretical models \citep[see, e.g.,][]{Pana1975}.  Under these assumption, we predict a maximum 
flux density for free-free emission of 0.2 mJy~beam$^{-1}$ at 342 GHz. This value is 20 times less than that measured, implying therefore 
that the central continuum emission must come mostly from dust grains.  

The brightness temperature of the central continuum emission is 5.3 K. This is much less than the physical dust temperature which is 
expected to be $\sim$ 1000-100 K in the range 0.5-10 AU. This suggests that the emission is either optically thin or it comes from 
an optically thick region much smaller than the synthesized beam. In order to have an idea of the extent of the inner disk, we 
performed a simple 
model assuming that the disk is optically thick, with an inclination of 70 degrees \citep{Mari2015}, and that the surface density 
scales with the radius as $r^{-1}$ and $\Sigma$(1 AU) = 3 g cm$^{-2}$. We calculated 
the temperature and emission at 342 GHz using the radiative code RADMC-3D \citep{Dull2012}. The method and the dust properties are 
detailed in section \ref{sec:detailed}. We found that a disk with a radius of 
2-3 AU reproduces the observed flux. If the inner disk is instead coplanar with the outer disk (i. e. with an inclination of 
27$^{\circ}$), its outer radius would be about 1 AU. These disk sizes are in agreement, or smaller for the coplanar case, with 
the tidal truncation expected by the observed companion at $\sim$ 13 AU \citep{Papa1977}.

Finally, the continuum map reveals a faint compact source  located at about 50 AU north of the center (PA $\sim$ 5\arcdeg). 
The flux density of this source, which is labeled in the Figure with the letter {\it c}, is 0.80 mJy beam$^{-1}$, or 6.9 times the noise 
level. Its brightness temperature is 3.5 K, suggesting either a very small source or optically thin emission. At such distances from the 
star, the temperature due to stellar irradiation can be estimated to $\sim$ 100 K and the observed flux would be consistent with an 
optically thick clump of dust of about 1.6 AU in size. If the emission is optically thin and the dust opacity is 2.9 cm$^{2}$ g$^{-1}$ 
as in \cite{Muto2015}, the measured flux density corresponds to a mass of $\sim$ 
1.5$\times$10$^{24}$ kg, or about a quarter of the Earth mass. Despite the low signal-to-noise ratio of the continuum emission, the 
observations suggest that source $c$ corresponds to a real structure in dust and gas. First, source $c$ is detected with a signal-to-noise 
of 6.9 while the second brightest peak emission inside the cavity has a signal-to-noise ratio of 3.5. Furthermore, the rms noise inside 
the cavity is the same as the noise in the outer parts of the map (130 $\mu$Jy beam$^{-1}$ instead of 115 $\mu$Jy beam$^{-1}$). 
Additionally, we verified that no negative components at the position of source $c$ exist in the map of the residual obtained at the end 
of the deconvolution process. This suggests that source $c$ is not a result of over-cleaning the dirty map. Finally, we have detected gas 
in non-Keplerian rotation at the position of source $c$ in the $^{13}$CO map (see Figure~\ref{Fig:spec-strea}). Deeper ALMA observations 
at higher angular resolution are required to assess the nature of source $c$.

No dust emission is detected somewhere else in the cavity, even adopting natural weighting for the imaging which delivers a sensitivity of 
65 $\mu$Jy beam$^{-1}$ and a beam of $0.36''\times0.31''$.  Using this weighting, the peak emission on the dust crescent 
is $\sim$ 240 mJy beam$^{-1}$. Considering we do not detect dust emission in the cavity above 3 $\sigma$ (195 $\mu$Jy beam$^{-1}$), 
the dust emission at millimeter wavelength has then dropped by a factor $>$ 1200 from the peak.

\subsubsection{CO integrated emission}
\label{sec:COint}

\begin{figure*}
  \includegraphics[angle=-90,width=\textwidth]{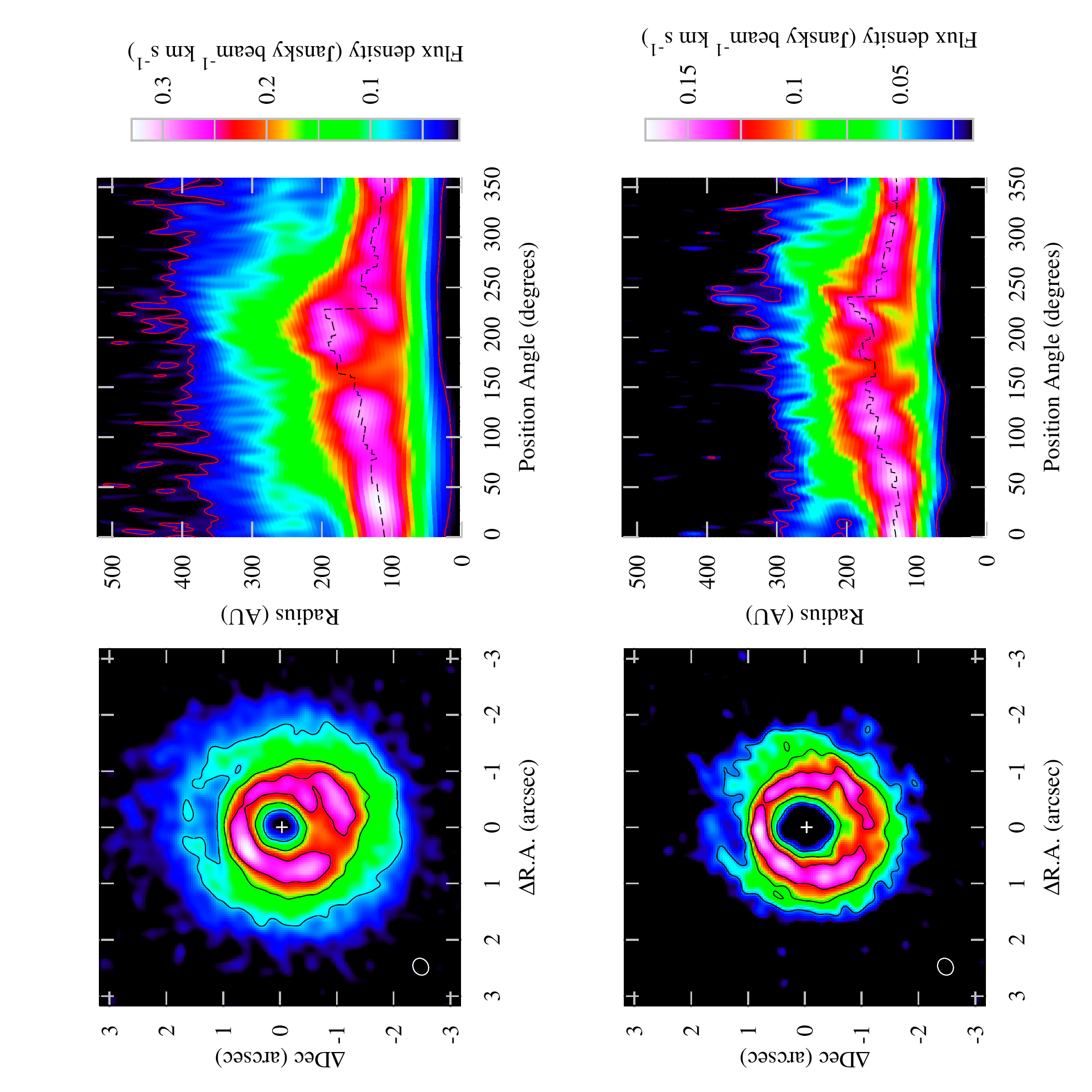}\\
  \caption{Top left: $^{13}$CO J=3-2 integrated intensity (Moment 0) map. The FWHM of the synthesized beam is 
  0.27\arcsec$\times$0.31\arcsec and is indicated by the white ellipse. The contours correspond to 25, 50 and 75 $\%$ 
  of the maximum emission. Top right: $^{13}$CO J=3-2 integrated intensity in function of radius and PA. The dashed line 
  represents the $^{13}$CO crest position in the disk. The red lines indicate the 3 $\sigma$ contours with $\sigma$ 
  $=$ 7.2 mJy beam$^{-1}$ km s$^{-1}$ for $^{13}$CO and 7.8 mJy beam$^{-1}$ km s$^{-1}$ for C$^{13}$O. Bottom 
  panels: Same as on the top but for C$^{18}$O J=3-2.} 
  \label{Fig:mom0CO}
\end{figure*}

Figure~\ref{Fig:mom0CO} displays spectrally integrated (Moment 0) maps of the $^{13}$CO J=3-2 and C$^{18}$O J=3-2 line 
emission, obtained after subtracting the dust continuum emission. The rms noise on these maps is 7.2 Jy beam$^{-1}$ km 
s$^{-1}$ for $^{13}$CO J=3-2 and 7.8 Jy beam$^{-1}$ km s$^{-1}$ for C$^{18}$O J=3-2.

% This pixel selection reduces the noise in the map and allows one to see the faint emission 
% arising from the outer disk. 

The $^{13}$CO and C$^{18}$O emission extends to about 400 AU and 300 AU from the center, respectively. By comparison, the 
outer radius of the dust emission is about 300 AU, similar to C$^{18}$O. In general, the azimuthal structure of the CO 
integrated emission differs substantially from that observed in the dust continuum. The CO emission is more azimuthally 
symmetric than the dust with only azimuthal variations by a factor of 1.5 for $^{13}$CO J=3-2 and 1.4 for C$^{18}$O J=3-2. 
The maximum is located at PA = 0-30$^\circ$ while the minimum is at PA $\sim$ 165$^\circ$.

Furthermore, a few structures are visible. First, the CO emission presents an apparent decrement at a distance of 170-220 AU from 
the star, on the north of the disk, corresponding with the continuum crescent. The interpretation of this feature is 
discussed in Section~\ref{sec:depletion}. Second, there are two local minima visible in both $^{13}$CO and C$^{18}$O integrated 
maps, on the south side at PA = 130$^\circ$-190$^\circ$ and, more faintly, on the north side at PA $\sim$ 330$^\circ$. They are 
at the same PA than the secondary minima observed in the dust emission, which was also previously seen in IR \citep{Aven2014}. 
The local decrease is more pronounced in $^{13}$CO, the most optically thick molecule, and suggests that the decrease is in temperature 
and affects predominantly the upper layers which are more sensitive to changes in stellar irradiation. This result supports the hypothesis 
of shadows cast by the inner circumprimary disk \citep{Mari2015}. Third, in the PA range 200-250\arcdeg\, the radial profile of the 
$^{13}$CO emission shows two peaks at about 115 AU and 200 AU from the star. This is characterized by the abrupt change of the 
crest radial position at PA = 230\arcdeg, visible on the top-right panel of Figure~\ref{Fig:mom0CO}. 

Both the $^{13}$CO and C$^{18}$O maps are characterized by much smaller central cavities than the cavity observed in the continuum. In 
C$^{18}$O, the 3 $\sigma$ contours indicates a cavity almost circular with radius between 60 and 70 AU. In $^{13}$CO, the cavity size 
varies with the position angle and measures  $\sim$ 15 AU at PA = 75\arcdeg\ and 35 AU at PA = 250\arcdeg. The minimum and 
maximum radius of the cavity in $^{13}$CO, dust, and, at a smaller degree, C$^{18}$O, are along the same azimuthal direction. 
As discussed in Section~\ref{sec:cavity}, this might suggest a common origin for their elliptical shape. 

%One possibility is tidal interactions with the stellar companion. \citep{Laco2016} separated 
%the possible orbits for the companion in two families where one has its periapsis and apoapsis at PA coherent with our observations. 
%Nevertheless, the presence of the stellar companion alone is surely not sufficient to explain the large size of the cavity and the 
%distant radial position of the crest. Finally, no CO emission is detected at the position of the central compact continuum source. 

\subsubsection{CO peak emission}
\label{sec:peak}

\begin{figure*}
  \includegraphics[angle=-90,width=\textwidth]{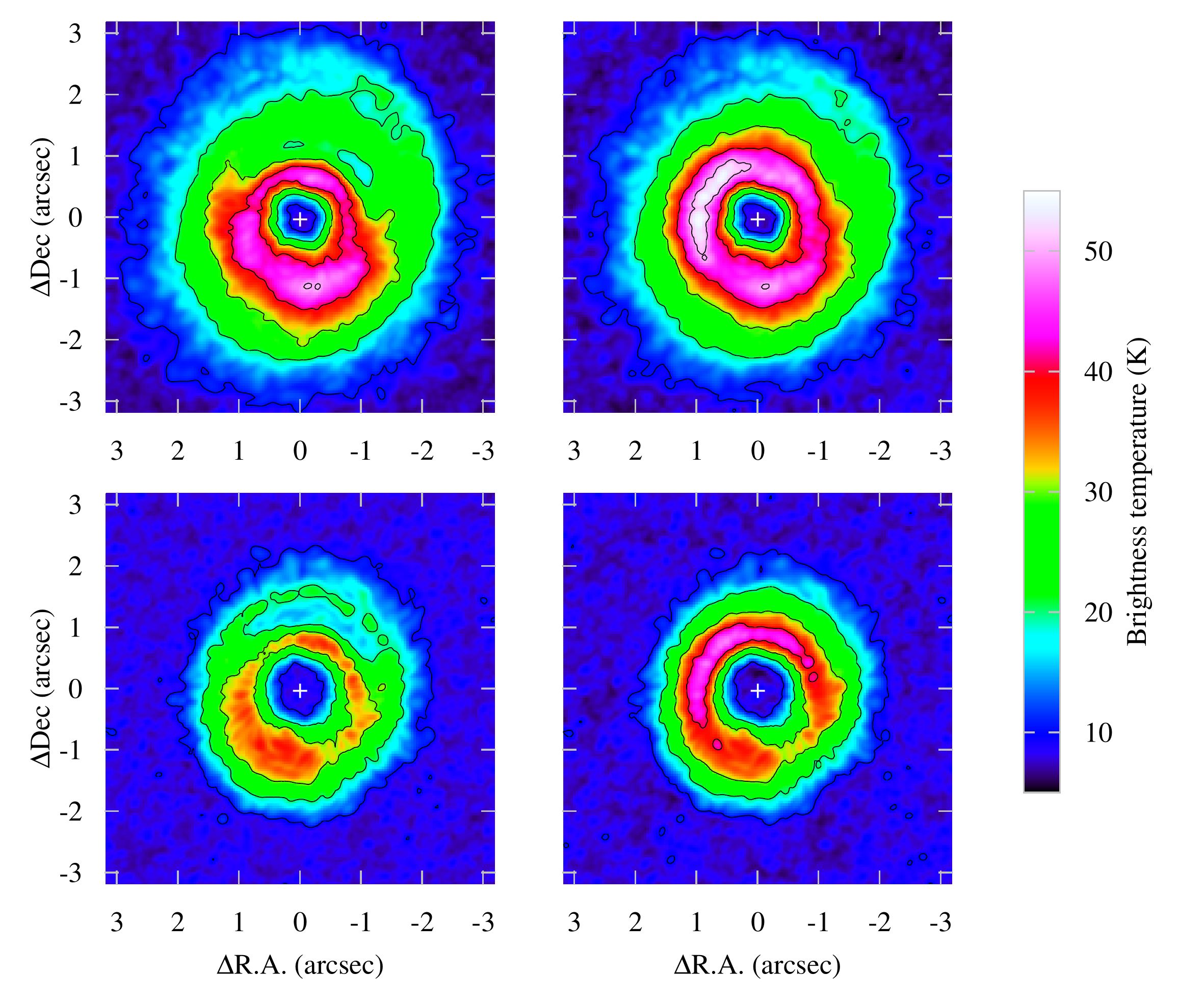}\\
  \caption{Peak emission maps in units of the brightness temperature for the $^{13}$CO (top) and C$^{18}$O J=3-2 (bottom) line 
  emission. The left and right panels show respectively maps with and without dust continuum subtraction. The contours are 
  plotted at 10, 20, 30, 40 and 50 K.}
  \label{Fig:mom8}
\end{figure*}

Figure~\ref{Fig:mom8} shows maps of the $^{13}$CO and C$^{18}$O peak emission. The main advantage 
of using the peak intensity is that it probes the most optically thick part of the line emission and constrains the gas 
temperature if the line center is optically thick. The figure shows the line intensity in units of brightness temperature 
defined as the temperature of a black body covering the area of the synthesized beam and emitting the observed flux density. 
For the conversion of Jy beam$^{-1}$ to Kelvin, we used the inverse of the Planck function:
\begin{equation}
  T = \frac{h\nu}{k} ~ \frac{1}{\mathrm{ln} \left( \frac{2h\nu^{3}}{F \Omega c^2}+1 \right)}
\end{equation}
with $h$ the Planck Constant, $k$ the Boltzmann Constant, the frequency $\nu$, the flux density F, and $\Omega$ the beam solid 
angle defined by $(\pi \theta_{max} \theta_{min})/(4\mathrm{ln}(2))$ where $\theta_{max}$ and $\theta_{min}$ are the major and minor FWHM 
of the synthesized beam.

% With a spectral resolution of 0.11 $\mathrm{km.s^{-1}}$ and characteristic emission detected on 1-2 $\mathrm{km.s^{-1}}$, 
% we have enough precision to measure the peak emission. One advantage of using the peak emission is that it can be converted 
% directly in brightness temperature using the inverse of the Planck function. 

In the left column of Figure~\ref{Fig:mom8}, we display the $^{13}$CO and C$^{18}$O peak emission calculated after subtracting 
the continuum emission. They present a brightness temperature decrease at the position of the continuum crescent which 
is more pronounced than in the integrated intensity maps. Taken at face value, this might suggest a local decrease in either line 
optical depth or gas temperature. 
On the right column of the figure, we show brightness temperature maps obtained without subtracting the continuum. As expected, 
the brightness temperature increases to account for the extra emission previously attributed to the continuum, and the 
decrement observed at the position of the continuum crescent disappears. In Section \ref{sec:depletion}, we claim that a peak 
intensity map obtained without subtracting the continuum emission is the best probe of the physical temperature of the gas if 
the emission at the line center is optically thick, as in the case of the $^{13}$CO J=3-2 emission in the outer disk of 
HD~142527 \citep{Muto2015}. The map shown in the top-right panel of Figure~\ref{Fig:mom8} therefore unveils the thermal 
structure of the $^{13}$CO J=3-2 emitting layer.

\begin{figure}
  \includegraphics[angle=-90,width=0.5\textwidth]{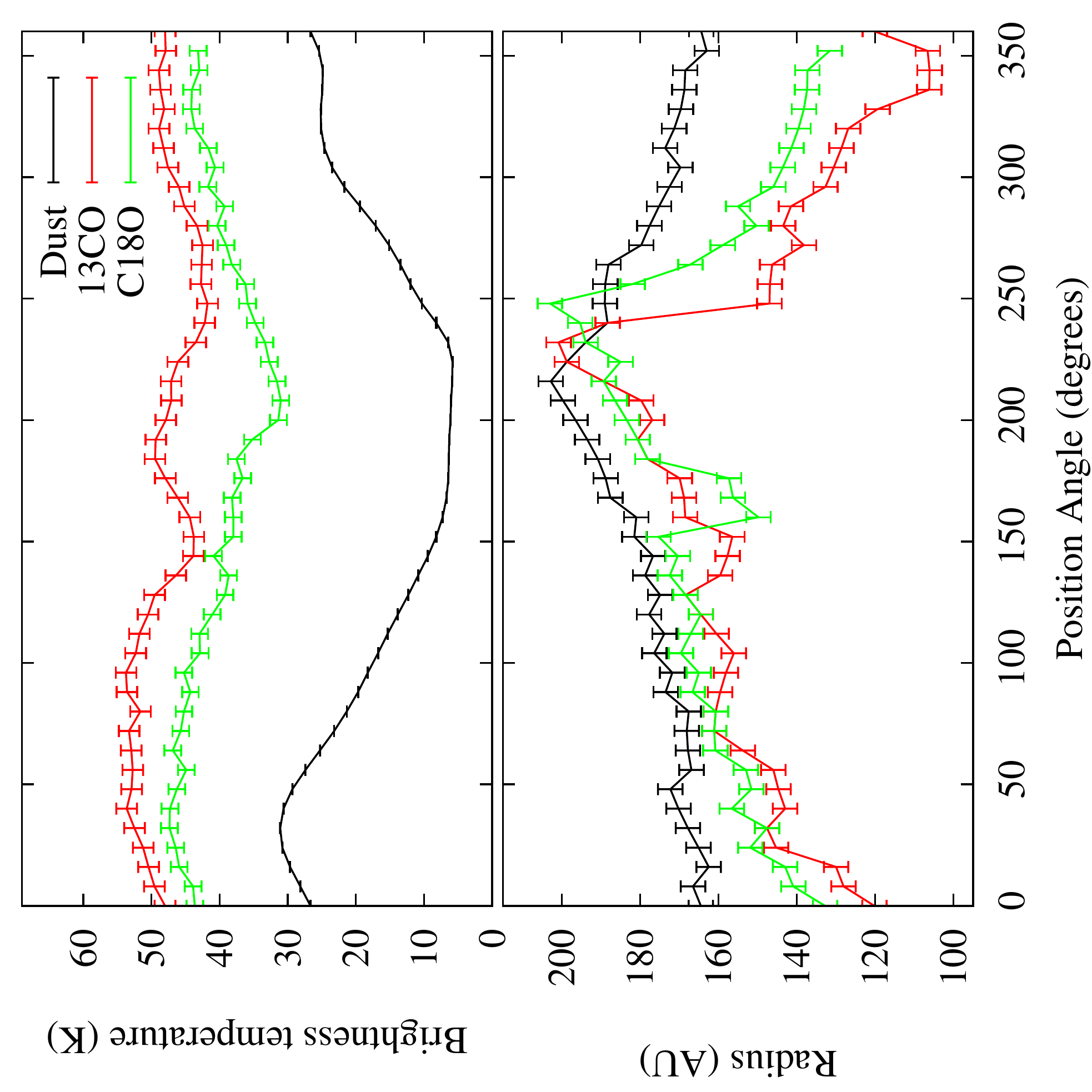}\\
  \caption{Top: Brightness temperature of the dust, $^{13}$CO J = 3-2 and C$^{18}$O J = 3-2 along the crest in function of the 
  position angle. Bottom: Radial position of the crest in function of the position angle.}
  \label{Fig:peak-max}
\end{figure}

Figure~\ref{Fig:peak-max} compares the peak emission coming from the crest of the two molecules with the dust continuum. The 
peak emission of the molecules is warmer than the dust emission with differences between dust and $^{13}$CO J=3-2 of $\sim$ 
20 K on the dust crescent and 40 K on the opposite side. The horseshoe structure is more pronounced in the dust with brightness 
temperature varying between 30 K on the dust crescent to 5 K on the opposite side but is also visible in C$^{18}$O, with temperature 
varying from 47 K to 31 K. 

The $^{13}$CO emission, more azimuthally symmetric, has a maximum brightness temperature of 54 K on the 
north-east part of the disk. In general, the eastern side of the disk is warmer than the western side, which is consistent 
with the east-west asymmetry observed in the mid-infrared thermal emission by \cite{Fuji2006}. The temperature difference probably 
comes from the disk inclination as the inner wall of the outer ring on the east side is exposed to the observer at a more face-on 
inclination. Also, the $^{13}$CO peak emission map shows a local minimum at PA = 130\arcdeg-180\arcdeg\ with a local decrease of 
$\sim$ 10 K, corresponding to the larger azimuthal minimum observed in the integrated map, and due probably to the inner disk shadow. 
We nevertheless do not see clearly this feature in C$^{18}$O or on the opposite side of the disk as in the integrated maps.

The bottom panel of Figure~\ref{Fig:peak-max} shows the radial position of the crests, which are located at smaller radii for the 
CO molecules than for the dust. $^{13}$CO, the most optically thick molecule, generally has its crest at smaller radii than 
C$^{18}$O. This suggests that the gas density at the cavity edge increases with radius on a spatial scale resolved by our observations. 
The smaller radius of the $^{13}$CO crest is then a result of the larger optical depth of this line. This is particularly visible at 
PA = 0\arcdeg-160\arcdeg\ and 250\arcdeg-360\arcdeg. Vice versa, at the azimuthal angles between 160\arcdeg\ and 250\arcdeg, the $^{13}$CO 
and C$^{18}$O crests are at the same distance from the star. As we will discuss later, in this region the emission of the lines are 
mostly optically thin and the crest likely indicates the peak of the gas surface density, instead of the transition between the 
optically thin and optically thick regions.  

Finally, the $^{13}$CO map in Figure~\ref{Fig:mom8} reveals a spiral structure starting form the west (at $\sim$ 2\arcsec) 
and propagating to the north of the disk up to 2.5\arcsec. This structure is consistent with the S1 spiral arm observed with 
ALMA in the $^{12}$CO J=3-2 and J=2-1 lines by \cite{Chri2014}. This spiral was also observed in near-infrared H and K bands, 
implying that it entrains both molecular gas and small grains \cite[see also][]{Aven2014}.

% By the observation of scattered 
%light in IR \citep{Aven2014}, the inner disk was modelled to have an inclination of 70 degrees compared to the outer disk 
%and for this reason only hides 2 azimuthal regions of the outer disk.

%For the first time, we were able to observe both, in the north side and in the south side, this effect at millimeter 
%wavelengths and then study the effect on the temperature. The disk temperature drops from 60K to 53K on the north side 
%and to 48 K on the south. Such discrepancy seems to imply that the inner disk is not azimuthally asymmetric, with more 
%optical and IR extinctions on the south side.  

\subsection{Disk dynamic}
\label{sec:dyn}

\begin{figure*}
  \includegraphics[angle=0,width=\textwidth]{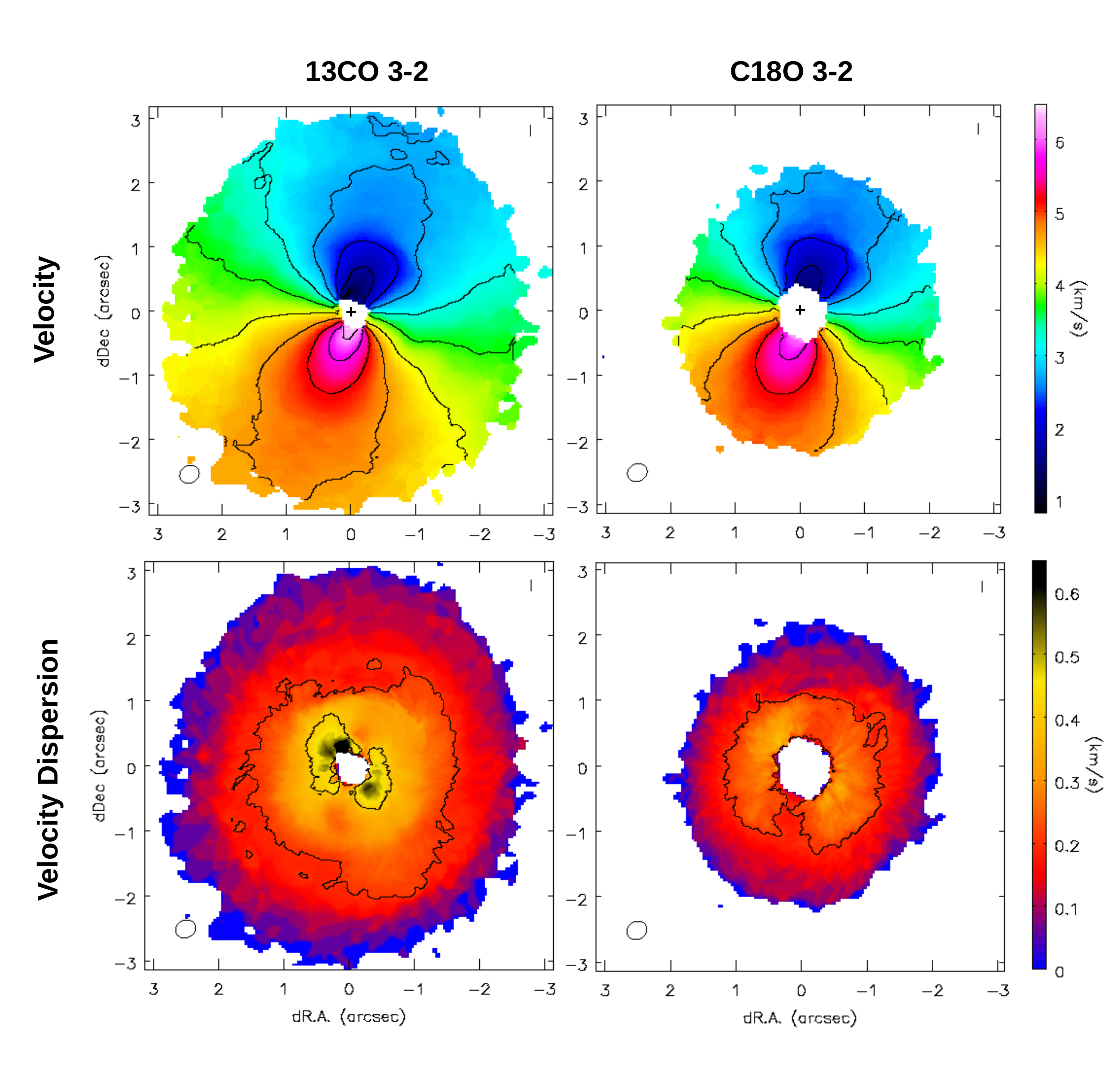}\\
  \caption{Top: Maps of the velocity centroid for the $^{13}$CO and C$^{18}$O J=3-2 lines 
  plotted between 0.8 to 6.5 $\mathrm{km s^{-1}}$. Contours start from 1.75 km s$^{-1}$ and are spaced by 
  0.5 $\mathrm{km s^{-1}}$. The systemic velocity is 3.75 km s$^{-1}$. Bottom: map of the velocity dispersion 
  plotted between 0 to 0.65 km s$^{-1}$. As a reference, contours have been drawn at 0.2, 0.4 and 0.6 km s$^{-1}$.}
  \label{Fig:mom1-2}
\end{figure*}

Maps of the velocity centroid of the $^{13}$CO and C$^{18}$O emission (moment I) are displayed on the top panels of 
Figure~\ref{Fig:mom1-2}. They were calculated by excluding pixels with emission below 5 $\sigma$  after subtracting 
the continuum emission. The rotation pattern of the disk is clearly visible. The north part of the disk moves toward 
us while the southern part moves away from us. From these maps, we measure a systemic velocity of 3.75 $\pm$ 0.05  
km s$^{-1}$ with the uncertainty given by half the size of the spectral bin, and a position angle of -19 
$\pm$ 1\arcdeg\ for the major axis of the disk.

\begin{figure}
  \includegraphics[angle=-90,width=0.5\textwidth]{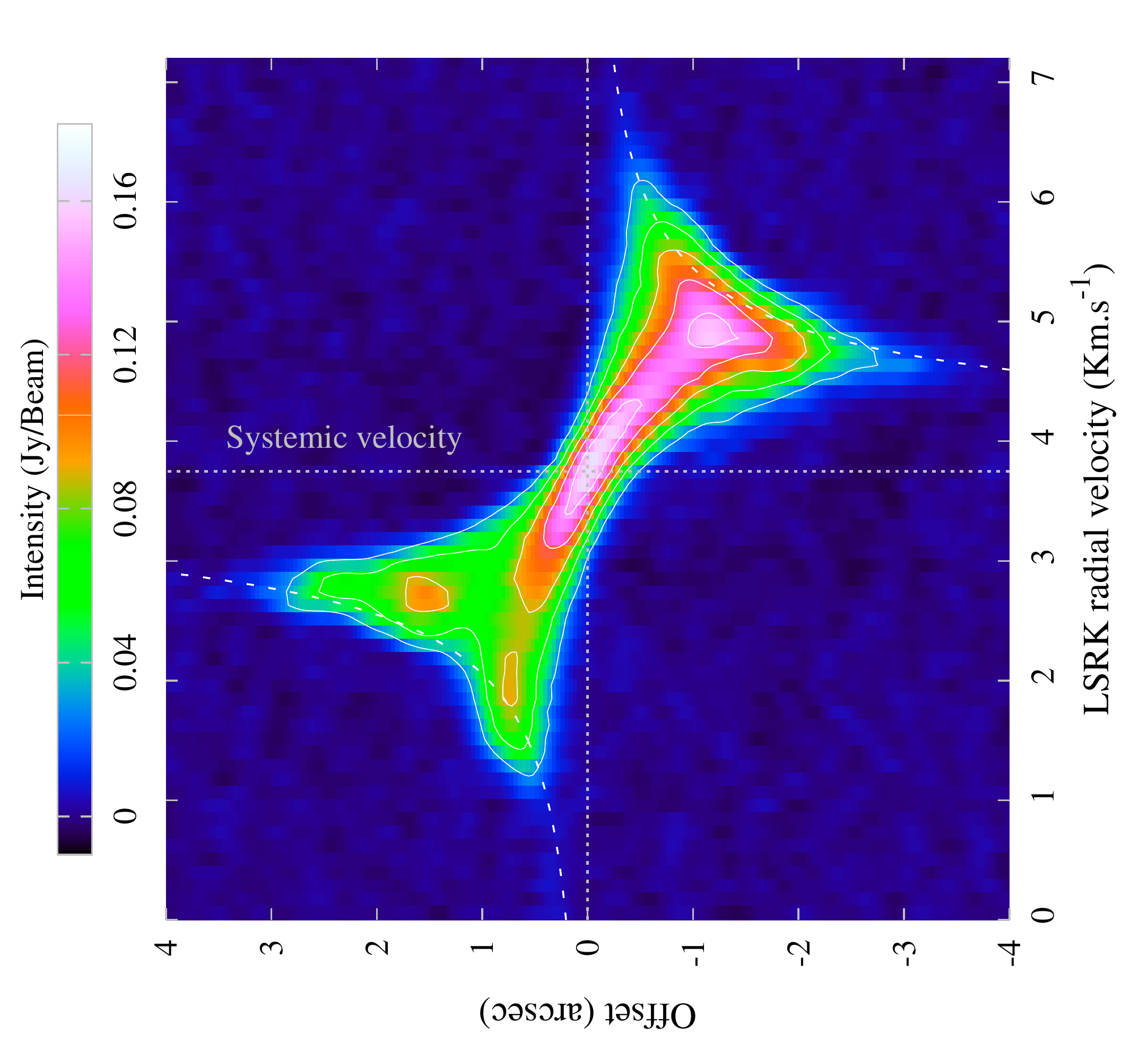}\\
  \caption{Position-Velocity diagram of $^{13}$CO J = 3-2  along the major axis (with PA = -19\arcdeg). The vertical dotted line 
  represents the systemic velocity at 3.75 km s$^{-1}$ and the horizontal 
  dotted line the zero distance to the minor axis. The white dashed lines show the Keplerian velocity for a star of 2.4 solar 
  masses and a disk inclination of 27\arcdeg. Contours are multiples of 5 of the noise level (i.e. 5.9 mJy beam$^{-1}$). }  
  \label{Fig:PV}
\end{figure}

Figure~\ref{Fig:PV} shows the Position-Velocity diagram along the major axis calculated from the $^{13}$CO line. We find 
that the rotation pattern is compatible with Keplerian rotation around a stellar mass of 2.4 M$_\odot$ and a disk inclination 
of about 27\arcdeg. The value of the stellar mass is in agreement with \cite{Verh2011}, given the new distance of 156 pc 
measured by Gaia \citep[][]{Gaiaa2016,Gaiab2016}. The disk inclination is in agreement with \cite{Fuka2013} and \cite{PerS2015}. 
The center of rotation derived from the moment I map and Position-Velocity diagram is located at the same position 
than the central continuum source discussed in the previous section. It is indicated by a black cross in the 
moment I map of $^{13}$CO and C$^{18}$O. The uncertainty on the position of the center of rotation is 0.05\arcsec. 

The bottom panels of Figure~\ref{Fig:mom1-2} show maps of the velocity dispersion of the $^{13}$CO and C$^{18}$O J = 3-2 
lines (moment II). The velocity dispersion decreases with the distance from the star as expected for an inclined Keplerian 
disk, and it is larger for the $^{13}$CO line, probably due to its higher opacity, which allows better detection of the 
line wings. The velocity dispersion measured in the $^{13}$CO line is lower than 0.65 km s$^{-1}$ across the entire 
disk with the exception of a small region at 0.3\arcsec\ north and 0.1\arcsec east of the center, where it reaches 1 
km~s$^{-1}$. This region coincides with the faint compact continuum source discussed above (source c). It is also in this region 
that \cite{Casa2013} claimed to have observed gas streamers in HCO+.

\begin{figure}
  \includegraphics[angle=0,width=0.5\textwidth]{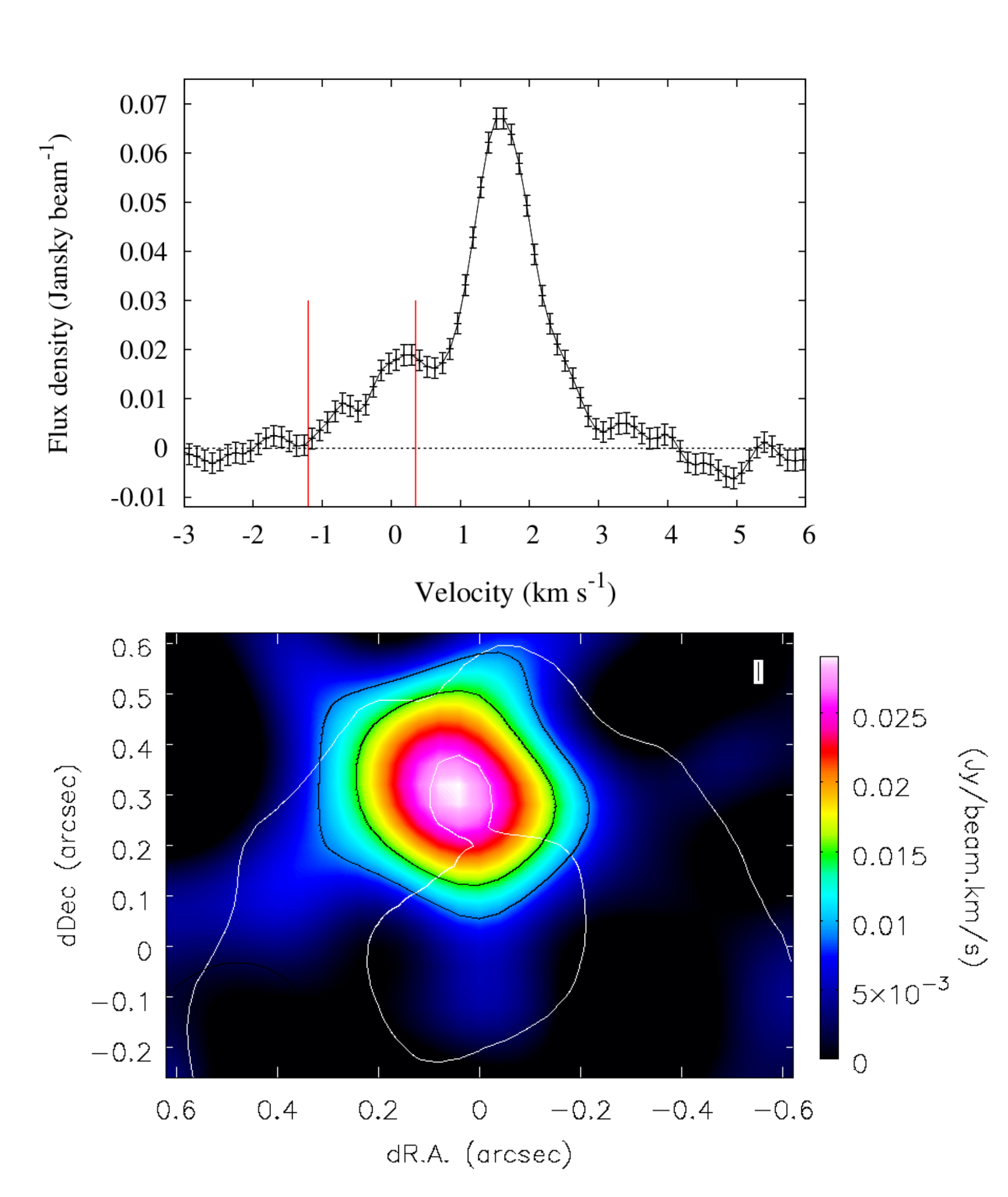}\\
  \caption{Top panel: Spectra of the emission corresponding to the disk region characterized by the highest velocity dispersion. 
  The spectrum is extracted at the position 0.03\arcsec-0.3\arcsec, using an ellipse with an area three times larger than the beam size. 
  The two vertical red lines show the interval in velocities used to calculate the integrated emission map coming from the shoulder of 
  the spectrum. bottom panel: Map of the integrated emission covering the velocity range between -1.150 and 0.300 km $s^{-1}$, 
  covering 14 channels. The black contours are the 3 and 5 $\sigma$ contours levels with $\sigma$ = 2.97 mJy beam$^{-1}$ km s$^{-1}$. 
  The white line is the 5 $\sigma$ contour of the dust emission with $\sigma$ = 115 $\mu$Jy beam$^{-1}$.}   
  \label{Fig:spec-strea}
\end{figure}

A $^{13}$CO J=3-2 line spectrum extracted from the region of large velocity dispersion is shown on the top panel of Figure 
\ref{Fig:spec-strea}. The spectrum presents a large shoulder in the velocity range -1.150 km s$^{-1}$ to 0.3 km s$^{-1}$. 
The integrated intensity map of the line emission in this velocity range is displayed on the bottom panel of Figure 
\ref{Fig:spec-strea}. Due to its association with the local dust continuum emission $c$, we argue that this emission might 
arise from a planetary disk or streamers accreted through a giant planet.

% The velocities of the brightest and faintest 
%peak with respect to the local standard of rest are 1.4 km s$^{-1}$ and 0.4 km s$^{-1}$, respectively, while their velocities 
%relative to the systemic velocities are -2.35~km~s$^{-1}$ and -3.35~km~s$^{-1}$, respectively. 

% These latter values can be 
%compared with the component of the Keplerian velocity along the line of sight expressed by $v_r = \sqrt{GM_\star/r} \sin{i} 
%\cos{\theta} \simeq$ 2 km~s$^{-1}$, where $\theta $\sim$ 0^{o}$ is the position angle of the region relative to the disk major 
%axis. 

% smoothed on 3 channels with the Hanning method.  We did not find such features in other parts of the disk.
% With the hypothesis of Keplerian rotation for both emission and taking into account the angle PA and the 
% inclination of the disk, the velocities of the 2 flows are -8 and -11.5 $\mathrm{km.s^{-1}}$ compared to the star.  

%Due to the beam dilution, both velocities could be in principle explained by Keplerian rotation. Indeed, the faintest emission, 
%which is barely visible in Figure~\ref{Fig:PV} at v= 0.4 km $s^{-1}$ and at an offset of 0.3\arcsec\, lays approximatively on 
%the Keplerian fit curve. However, the shoulder and especially the double peak spectral line is not consistent with pure Keplerian 
%rotation. 

\section{Analysis}
\label{sec:analysis}
\subsection{On the anti-correlation between dust and molecular line emission}    
\label{sec:depletion}

A decrease in the molecular line intensity from the disk region corresponding to the dust continuum crescent 
has been observed in several tracers including $^{12}$CO J=2-1, $^{13}$CO J=3-2, C$^{18}$O J=3-2, HCN J=4-3, 
CS J=7-6 \citep{PerS2015,vdP2014}. A similar anticorrelation between dust and gas emission was found in 
other disks characterized by continuum crescents, such as, for example, IRS~48 \citep{Brud2014}. 
\cite{Casa2015-2} detected a concentration of cm-size grains within the HD~142527 crescent and suggested that the 
decrease in CO emission might result from a depletion of CO molecules, caused by freeze-out onto these large and 
cold grains. However, our observations suggest a gas temperature of about 40 K in this region, 
 which is a factor of 2 higher than the CO freeze-out temperature. 

Alternatively, \cite{Muto2015} modeled the CO observations of the disk around HD~142527 and found that the observed decrease 
in CO intensity might be due to the absorption of the molecular line by the optically thick dust, and therefore does 
not trace a region of low CO density. We argue that this cannot be the main reason for strongly optically thick molecules 
as $^{12}$CO, $^{13}$CO and probably C$^{18}$O because the optical depth of the lines are very large compared to the optical 
depth of the continuum. Indeed, assuming fractional abundances of 6$\times 10^{-5}$, 9$\times 10^{-7}$ and 1.35$\times 
10^{-7}$ for CO, $^{13}$CO and C$^{18}$O respectively \citep{Qi2011}, a gas-to-dust ratio of 100, a typical dust 
opacity of 2.9 cm$^2$ g$^{-1}$ at the lines frequencies, and a gas temperature of 40 K, we calculate that 
the emission at the center of the $^{12}$CO, $^{13}$CO and C$^{18}$O J=3-2 lines is about 1.1$\times$10$^{4}$, 170 
and 25 times more optically thick than the dust, respectively. Therefore, the absorption of the line emission by the 
dust at the lines center is negligible. 

We propose instead that the anti-correlation between dust and gas emission is an artifact resulting from the 
technique used to calculate and subtract the dust continuum contribution from the original CO+dust maps. 
The continuum emission is usually estimated by performing a linear extrapolation of its emission 
in line-free channels. However, when the continuum is absorbed by the molecule at the line frequency, this leads to 
overestimate the dust contribution and then underestimating the line emission. This effect is important 
if the line emission is optically thick (such that it absorbs the continuum) and comparable in intensity to the 
 line-free continuum emission (such that the effect of the subtraction  is the greatest). In practice, we argue that the 
 continuum-subtracted observations of dense environments might significantly  underestimate line emission. 

\begin{figure*}
  \label{Fig:dust-sub}
  \includegraphics[angle=-90,width=0.96\textwidth]{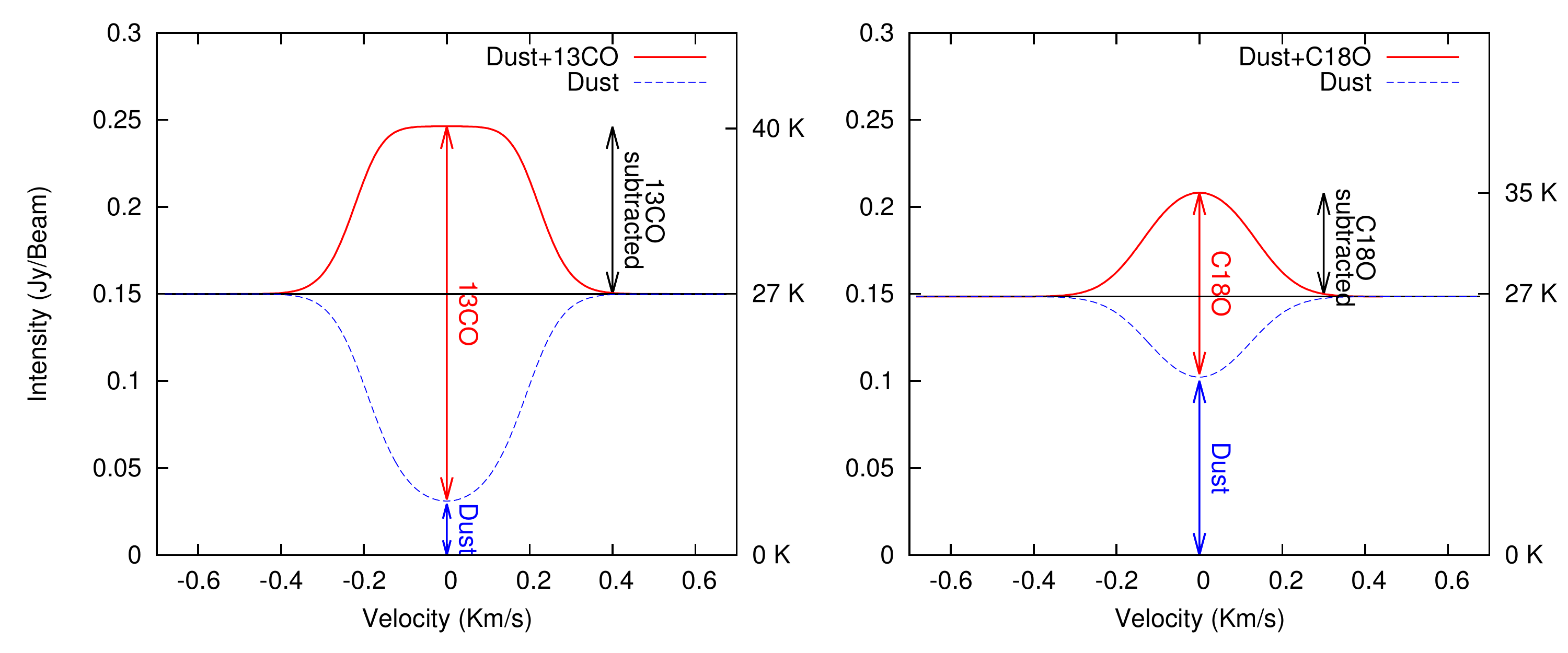}\\
  \caption{Effect of continuum subtraction on $^{13}$CO and C$^{18}$O J=3-2 lines intensity in the dust crescent. The 
  blue dashed line shows the emission from the dust continuum only, while the solid red line corresponds to the 
  continuum+line emission. Inside the line, the dust continuum emission drops due to the absorption by CO molecules. 
  The red and blue vertical arrows show the line and continuum contribution, respectively, to the total 
  emission at the lines center. The black vertical arrows indicate the CO peak emission after the usual dust subtraction.}  
  \label{Fig: dust-sub}
\end{figure*}

To quantitatively explain how this process works, we show in Figure \ref{Fig: dust-sub} examples spectra of 
the dust, $^{13}$CO J=3-2 and C$^{18}$O J=3-2 emission calculated for local conditions encountered in the dust crescent 
in the HD~142527 disk. We calculate the spectrum of the continuum outside and inside the line, assuming a vertically isothermal 
slab of gas at a temperature of 40 K, which corresponds to the brightness temperature at the dust crescent derived from the 
dust+$^{13}$CO peak emission map. Dust and gas are well mixed and the slab 
is seen face on. The continuum emission at the line frequencies, shown in blue dashed lines, is largely absorbed by CO 
molecules. In this case, a linear fit of the continuum based on the level measured in the line-free channels, represented 
by the horizontal black lines, is not adequate and will lead to an underestimate of the CO emission. This effect is the 
strongest at the center of the line, where the line opacity is the highest. This explains why the reduction in CO intensity 
is larger in the peak emission maps than in the integrated emission maps.   

In order to reproduce the observed peak emission of $^{13}$CO and C$^{18}$O, we had to adopt a gas-to-dust ratio of 4. This is 
a very small ratio compared to the standard value of 100 and might result from the high dust concentration. This corresponds to a 
dust surface density of 0.34 g~cm$^{-2}$ and a gas density of 1.4 g~cm$^{-2}$. The optical depths are 0.94, 6.5 and 0.97 for the 
dust, $^{13}$CO and C$^{18}$O J= 3-2, respectively. C$^{18}$O and the dust have similar optical depths and we observe, on 
the right panel of Figure~\ref{Fig: dust-sub}, that the dust and C$^{18}$O contributions to the total emission are roughly equivalent. 
In this case, the observed decrease of C$^{18}$O emission at the position of the dust crescent is due both to continuum 
subtraction (which affect the the optically thick center of the line) and to line absorption by the dust grains (which affects the optically 
thin wings of the line). 
On the contrary, the absorbed continuum emission at the center of the  $^{13}$CO is very little and the observed decrement 
in $^{13}$CO can be explained as the result of the continuum subtraction process. A more precise calculation of the gas 
and dust structure including a realistic vertical temperature profile is presented in Section \ref{sec:detailed}. 

In conclusion, we argue that the incorrect evaluation and subtraction of continuum emission is the main cause of 
the apparent anti-correlation between dust and $^{13}$CO emission observed in Figure~2 and also for $^{12}$CO in \cite{Pere2014}. 
Nevertheless, for less abundant molecules as C$^{18}$O, the line absorption by the dust grain becomes also important. The 
exact contribution of both processes will depend on the dust and molecular relative opacities as well as on their 
vertical distribution. In all cases, strong effects will only be visible in regions where dust and gas are optically 
thick. Direct interpretation of the lines emission in areas with large dust opacities must be done with caution.

\subsection{Derivation of the gas and dust surface density}
\label{sec:detailed}

A detailed analysis of the structure of the disk around HD~142527 must take into account the vertical gradient of the temperature and 
uses radial profiles to describe the gas and dust surface density. This can be obtained by performing radiative transfer followed by  
ray tracing. A global 3D analysis of the data is out of the scope of this paper due to its complexity and large computational time.  
Instead, we model the radial profiles of the dust and gas peak and integrated emission measured 
along the two azimuthal directions intersecting the maximum (PA $\sim$ 21\arcdeg) and minimum (PA $\sim$ 221\arcdeg) 
of the dust continuum emission. A similar procedure was adopted in modeling previous observations at an angular resolution 
of $\sim$ 0.45\arcsec\ by \citet[][hereafter M15]{Muto2015}. 

\subsubsection{Comparison with previous models}

\begin{figure*}
  \label{Fig:radcut}
  \includegraphics[angle=270,width=1.0\textwidth]{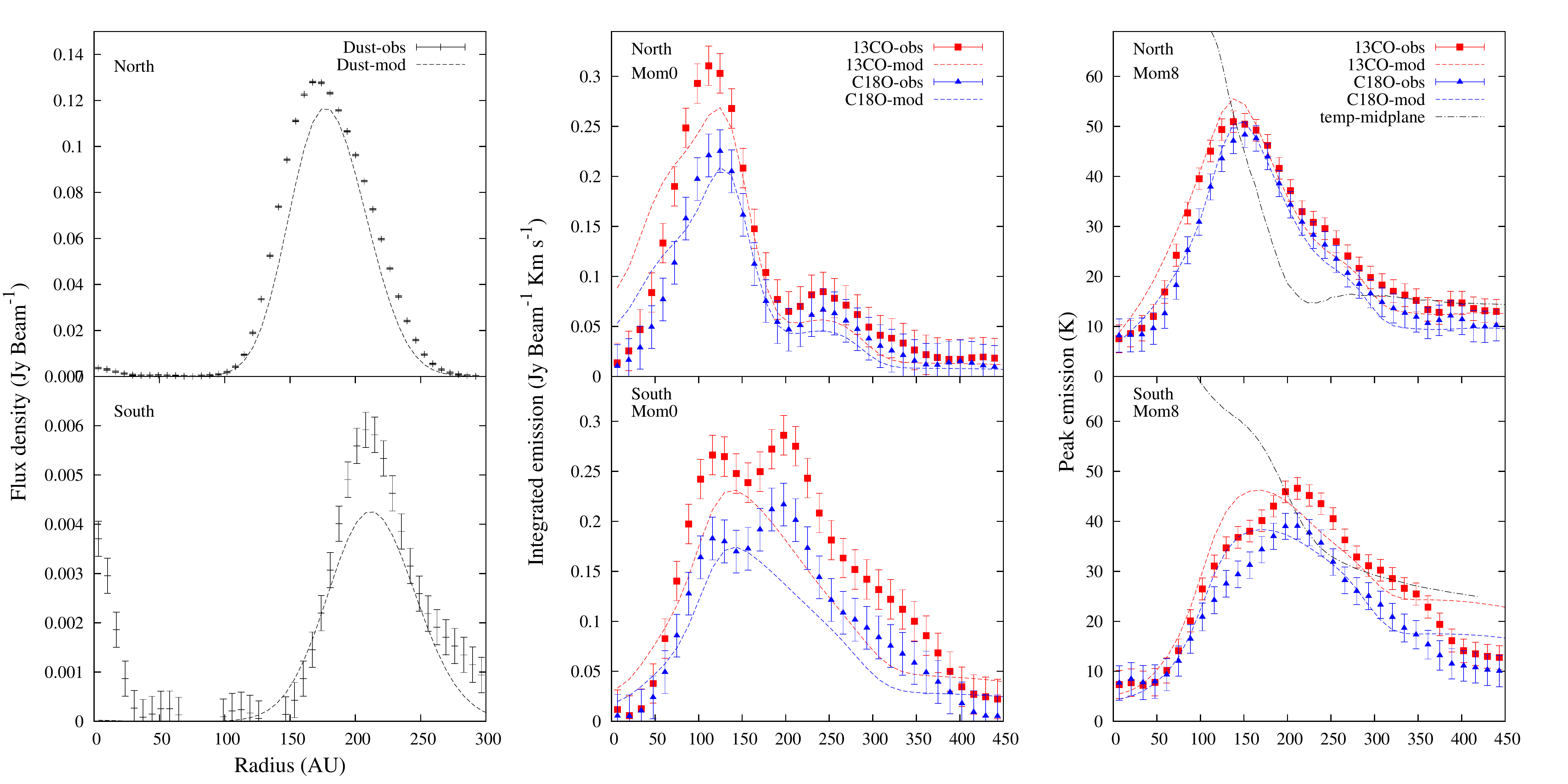}\\
  \caption{Radial cut along the north and south profiles at PA = 16\arcdeg-26\arcdeg and 216\arcdeg-226\arcdeg, respectively. Left: 
  Dust emission. Center: $^{13}$CO and C$^{18}$O J=3-2 integrated emission. Right: $^{13}$CO and C$^{18}$O J=3-2 peak emission. The 
  observations are represented with their error bars using red squares for $^{13}$CO and blue triangles for C$^{18}$O. The dashed lines 
  are the M15 models. On the right panels, the black dot-dashed line indicates the temperature on the disk midplane given by the model.}
  \label{fig:opticalD0}
\end{figure*}

As a first step, we compare our observations with the M15 model, rescaled to the distance of 156 pc, which is defined as follow. 
The dust surface density is expressed by a Gaussian law 
\begin{equation}
  \Sigma_d(r) = \Sigma_{d,0} ~ \mathrm{exp}\left[ -\left( \frac{r-r_d}{w_d} \right)^2 \right].
\end{equation}
In the north direction, defined by position angles between 16\arcdeg\ and 26\arcdeg, the dust surface density has 
$\Sigma_{d,0}$ = 0.6 g cm$^{-2}$, $r_d$ = 193 AU and $w_d = 30$ AU. In the south direction, defined by position angles 
between 216\arcdeg\ and 226\arcdeg, the dust surface density has $\Sigma_{d,0}$ = 8.45 $\times$ 10$^{-3}$ g cm$^{-2}$, 
$r_d$ = 218 AU and $w_d = 38$ AU. 

The disk temperature is calculated with the radiative code RADMC-3D \citep{Dull2012} assuming thermal equilibrium. The 
stellar irradiation comes from the central Herbig star of 2.4 solar masses, a luminosity of about 23 $L_\odot$ and a 
surface temperature of 6250 K. In this analysis, we neglect the irradiation coming from the stellar companion of $\sim$ 
0.25 M$_\odot$. %We also note that the position angles we model are not below the shadows of the circumprimary disk. 
The dust and the gas have the same vertical distribution, which is assumed to be in hydrostatic equilibrium with the 
midplane temperature. A few iterations between the vertical distribution and the disk temperature are needed to find 
a stable solution. 

The dust grain properties are chosen to have a dust opacity of 2.9 g cm$^{-2}$ at 0.88 mm, as in M15. 
This value corresponds to the mean opacity of porous grains made of silicates (25.7\%), carbon (18\%) and 
water ice (56.3\%), with sizes $a$ between 0.05 $\mu$m to 1 mm, and a  grain size distribution 
proportional to $a^{-3.5}$. The observations are then simulated with RADMC-3D, performing ray-tracing on the 
disk with the same inclination and position angle than HD~142527 and smoothed via CASA at the same angular resolution 
than our observation.

The left panels of Figure~\ref{fig:opticalD0} show the comparison between M15 dust model and our observations. Overall, 
the model reproduces well the radial position of the peak but underestimates the observed flux. Differently 
from M15, we do not include dust scattering in calculating the disk emission. We made this choice because millimeter-wave 
dust scattering opacities strongly depend on the composition and structure of (sub)-millimeter particles which are largely 
uncertain (see Appendix \ref{app:A}).

The gas surface density in M15 is parameterized using a broken power law
\begin{equation} \label{eq:gas-Muto}
  \Sigma_g(r)=\left\{
     \begin{array}{ll}
         f_{in} ~ \sigma_g(r_c) ~ \left(\frac{r}{r_c}\right) & (r < r_c)  \\
         \sigma_g(r)                                     & (r_c < r < r_{out})  \\
         f_{out} ~ \sigma_g(r_{out})                       & (r > r_{out})  \\
  \end{array}
  \right.
\end{equation}
with $\sigma_g$(r) = $\sigma_{g,0} \left( \frac{r}{200 AU} \right)^{-1}$, $r_c$ the cavity radius, and $r_{out}$ the outer edge 
of the dense disk. At these radii, the gas surface density decreases sharply by a factor $f_{in}$ and $f_{out}$, respectively. 
Inside the cavity, for $r < r_c$, the surface density increases linearly with radius, while for $r > r_{out}$, the gas surface 
density is constant. The values for the north profile are $\sigma_{g,0}$ = 0.94 g cm$^{-2}$, $r_c$ = 123 AU, $r_{out}$ = 279 AU, 
$f_{in}$ = 1./8. and $f_{out}$ = 0.01. On the south profile, the values are $\sigma_{g,0}$ = 0.26 g cm$^{-2}$, $r_c$ = 111 AU, $r_{out}$ = 
279 AU, $f_{in}$ = 1./8. and $f_{out}$ = 0.1. The fractional abundances of $^{13}$CO and C$^{18}$O are 9$\times$10$^{-7}$ and 1.35$\times$10$^{-7}$, 
respectively. Finally, as in M15, the line profile is calculated accounting for the thermal broadening 
and Keplerian rotation but does not include any turbulence.

Figure~\ref{fig:opticalD0} shows the comparison between the M15 model and our observations. Dust emission is shown on the left panels, while 
the spectrally integrated and peak CO emission are plotted on the central and right panels, respectively. In the right panel, we also 
plot the profile of the disk temperature in the disk midplane. As discussed below, the comparison between peak 
brightness temperature and physical disk temperature provides a rough idea of the optical depth for the CO line emission. 
In the north direction, the M15 model provides a good fit for both the integrated emission and 
the peak emission,  even if the emission in the cavity is a little overestimated. The model also correctly reproduces the 
decrease in integrated line emission observed at the position of the dust crescent between 130 AU and 240 AU from the star. 

However, the M15 model fails to reproduce the observations along the south direction. In particular, it under-predicts the radius 
of the peak of CO intensity. In our observations, the highest peak is located at an orbital radius of about 200 AU, 
while the model predicts the peak to be located at about 150 AU from the star. Additionally, the integrated CO emission 
presents a double peaked structure which is not captured by the model.

\subsubsection{Probing the dust trap hypothesis}

\begin{figure*}
  \label{Fig:radcut}
  \includegraphics[angle=270,width=1.0\textwidth]{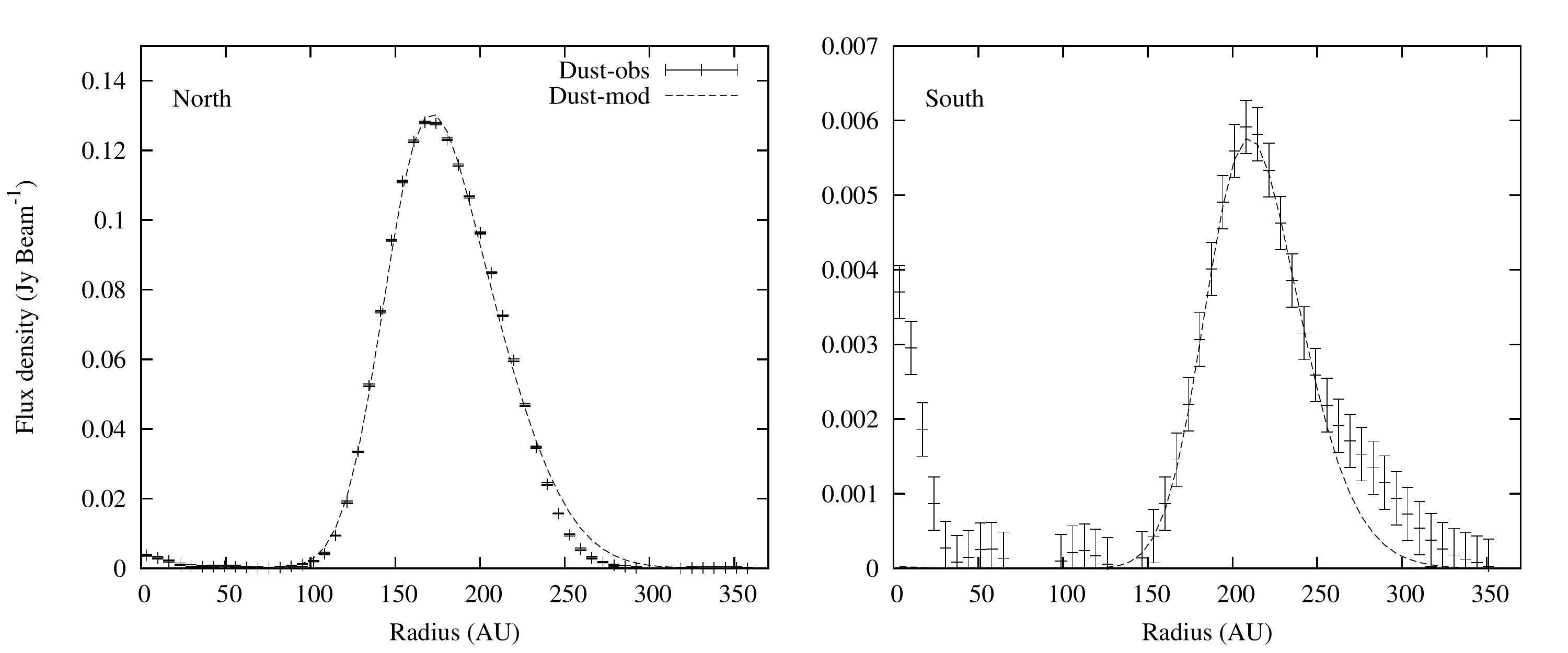}\\
  \caption{As for Fig. \ref{fig:opticalD0} but only for the dust. For modeling the surface density, we use a Gaussian distribution 
  with a different value for the inner and the outer width.}
  \label{fig:opticalD1}
\end{figure*}

Whereas the M15 model provides an acceptable fit to our observations, at least along the north direction, the assumed profile 
for the gas surface density (Equation~\ref{eq:gas-Muto}) is not physically motivated. In particular, it is unclear why the dust and gas surface 
densities should have different shapes and what might be the origin of sharp edges in the gas distribution at $r_c$ and $r_{out}$.  
In this section, we adopt a different approach and model the gas and dust radial distributions using Gaussian functions centered 
at the same radius r$_{r}$ and characterized by different widths. This parametrization mimics the effect of the trapping of dust grains 
by a Gaussian gas surface density profile peaking at  $r_r$ (see, e.g., \citealt{Pini2012}). Indeed, dust trapping causes the 
dust distribution to be more peaked toward the gas density maximum. To add an additional degree of freedom, 
we allow the width of Gaussian distribution for $r < r_r$ to be different from the width for $r>r_r$.  The dust distribution is therefore 
defined by 
\begin{equation}
  \Sigma_d(r)=\left\{
     \begin{array}{ll}
         \Sigma_{d,0}~\mathrm{exp}\left[ -\left( \frac{r-r_r}{w_{d,in}} \right)^2 \right] & (r < r_r)  \\
         & \\
         \Sigma_{d,0}~\mathrm{exp}\left[ -\left( \frac{r-r_r}{w_{d,out}} \right)^2 \right] & (r > r_r)  \\
  \end{array}
  \right.
  \label{eq:dustpr}
\end{equation}
The gas distribution is described by the same equation but with different values $\Sigma_{g,0}$, $w_{g,in}$ and $w_{g,out}$. 
Using this parameterization for the gas and dust surface density, we repeat the calculation of the disk temperature and 
continuum and line emission as discussed above. However, differently from M15 model, we calculate the line emission 
adding a constant turbulence velocity of 0.2 km s$^{-1}$ across the disk. 

In Figure~\ref{fig:opticalD1}, we present our best fit for the dust emission. The northern dust density profile has a maximum density 
of 0.65 g cm$^{-2}$ at $r_r =185$ AU.  The inner width is 28 AU while the outer width is 50 AU. The southern profile has a maximum 
surface density of 0.012 g cm$^{-2}$ at $r_r = 205$ AU and inner and outer widths of 17 AU and 43 AU, respectively.  This model provides 
a good fit to the observations between 25-230 AU. Instead, the dust distribution at larger distances seems to deviate from a simple Gaussian 
profile. Within 25 AU, the dust emission traces the compact source discussed above, which is not included in our model.

\begin{figure*}
  \label{Fig:radcut}
  \includegraphics[angle=270,width=1.0\textwidth]{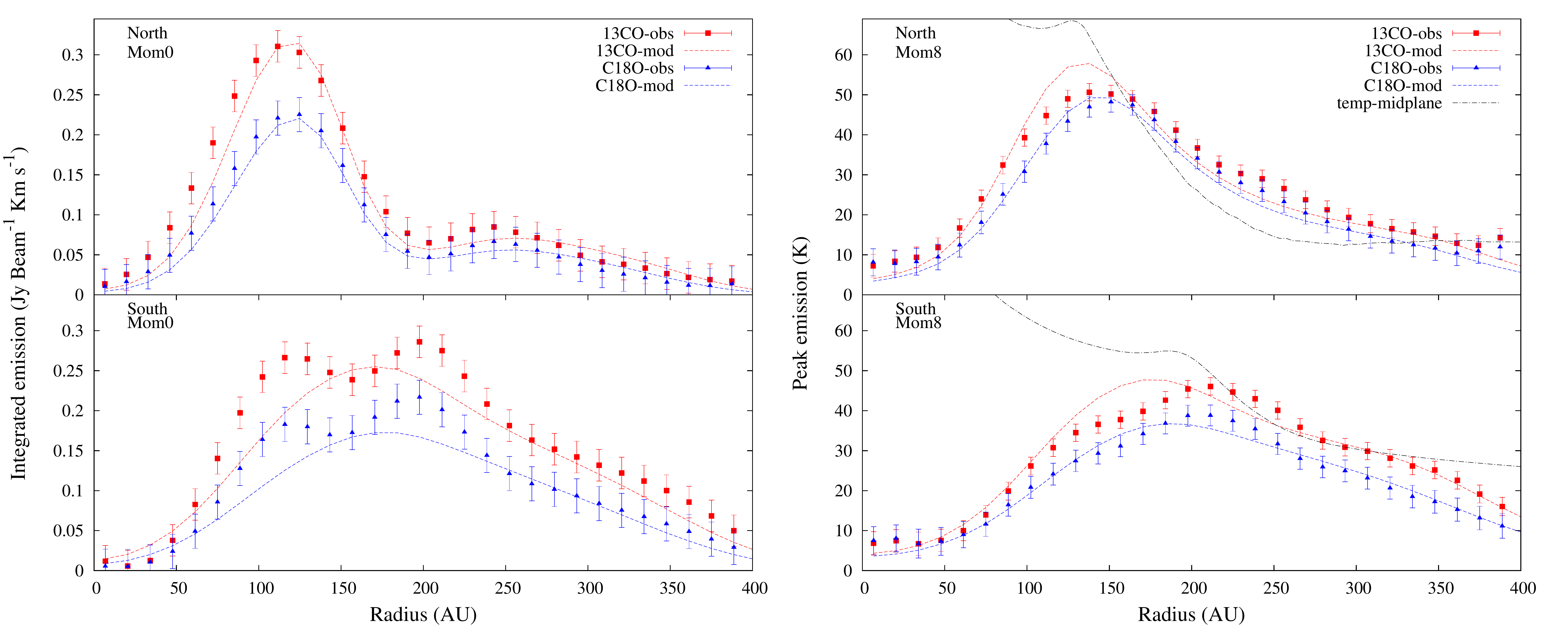}\\
  \caption{As for Fig. \ref{fig:opticalD0} but only for the gas. For modeling the surface density, we use a Gaussian distribution 
  with a different value for the inner and the outer width.}
  \label{fig:opticalD2}
\end{figure*}

The best fit model for the CO emission is shown in Figure~\ref{fig:opticalD2}. Along the north direction, the gas surface 
density is $\sim$ 1.1 g cm$^{-2}$ at $r_r = 185$ AU and has widths of 60 AU and 85 AU inside and outside the peak, respectively. 
On the south profile, the gas surface density is 0.24 g cm$^{-2}$ at $r_r = 205$ AU and the widths are 90 AU for the inner side and 
100 AU for the outer side. Along the north direction, our model well reproduces all the main features of 
the observations. However, there is still a discrepancy on the south profile, where the model fails in reproducing the secondary 
maximum on the integrated emission and the shoulder on the peak emission.

\begin{figure*}
  \includegraphics[angle=270,width=1.0\textwidth]{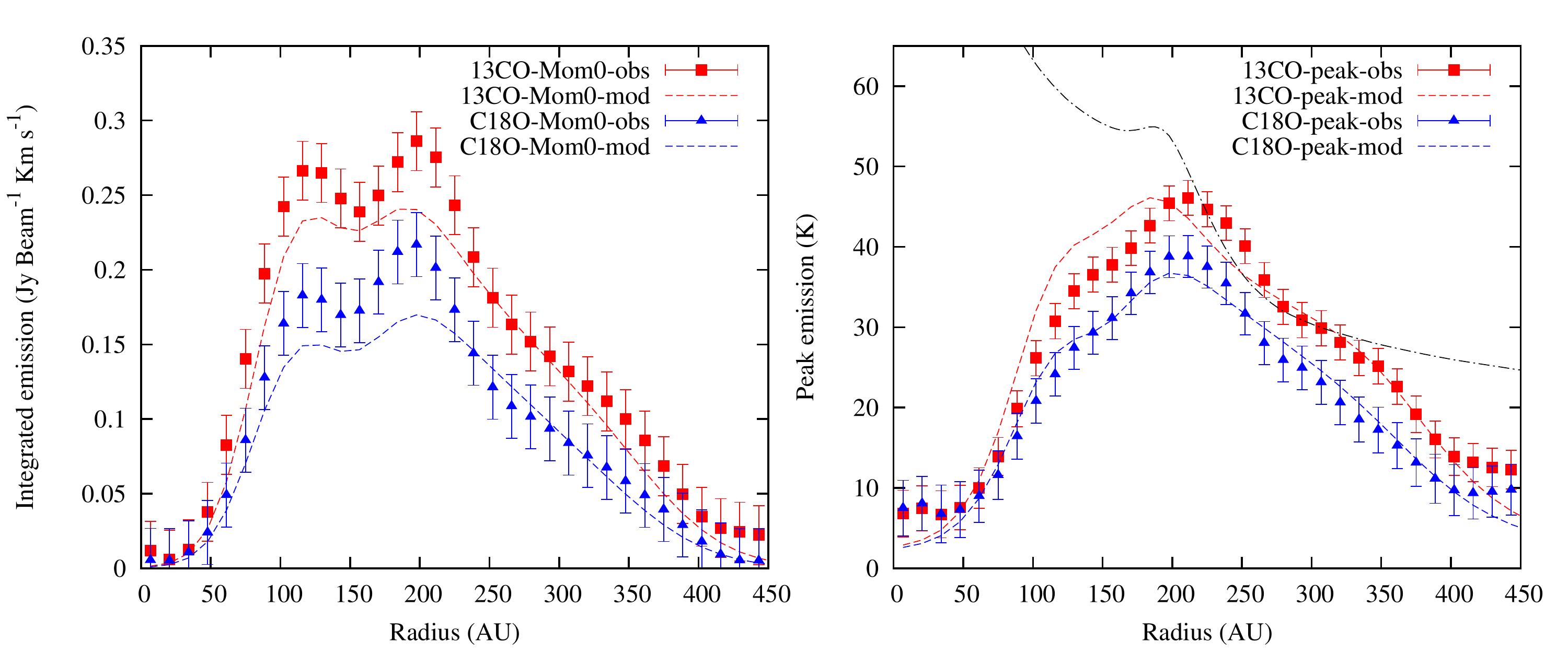}\\
  \caption{As for Fig. \ref{fig:opticalD0} but only for the gas on the south profile. The gas surface density is now modeled by using 2 
  Gaussians with different inner and outer widths.}
  \label{Fig:opticalD3}
\end{figure*}

The comparison between the physical disk temperature and the brightness temperature of the line emission (see the right panel of 
Figure~\ref{fig:opticalD2}) indicates that both lines are optically thin for $r \leq 200$ AU, as the brightness 
temperature of the 
emission is inferior to the disk midplane temperature. The secondary local maximum observed in the south profile at  $r  = 115$ AU 
might then come from a local enhancement in the gas density. In order to reproduce it, we added a second Gaussian in 
the description of the gas surface density for the south profile, such that the gas surface density is now defined as 
\begin{equation}
\sigma_g(r) = \Sigma_g(r) + \Sigma'_g(r)
\end{equation}
where $\Sigma_g(r)$ is as in Equation 4, and 
\begin{equation}
   \Sigma'_g(r)= \Sigma'_{g,0}~\mathrm{exp}\left[ -\left( \frac{r-r'_r}{w'_{g}} \right)^2 \right].
\end{equation}
 
\begin{figure}
  \includegraphics[angle=270,width=0.48\textwidth]{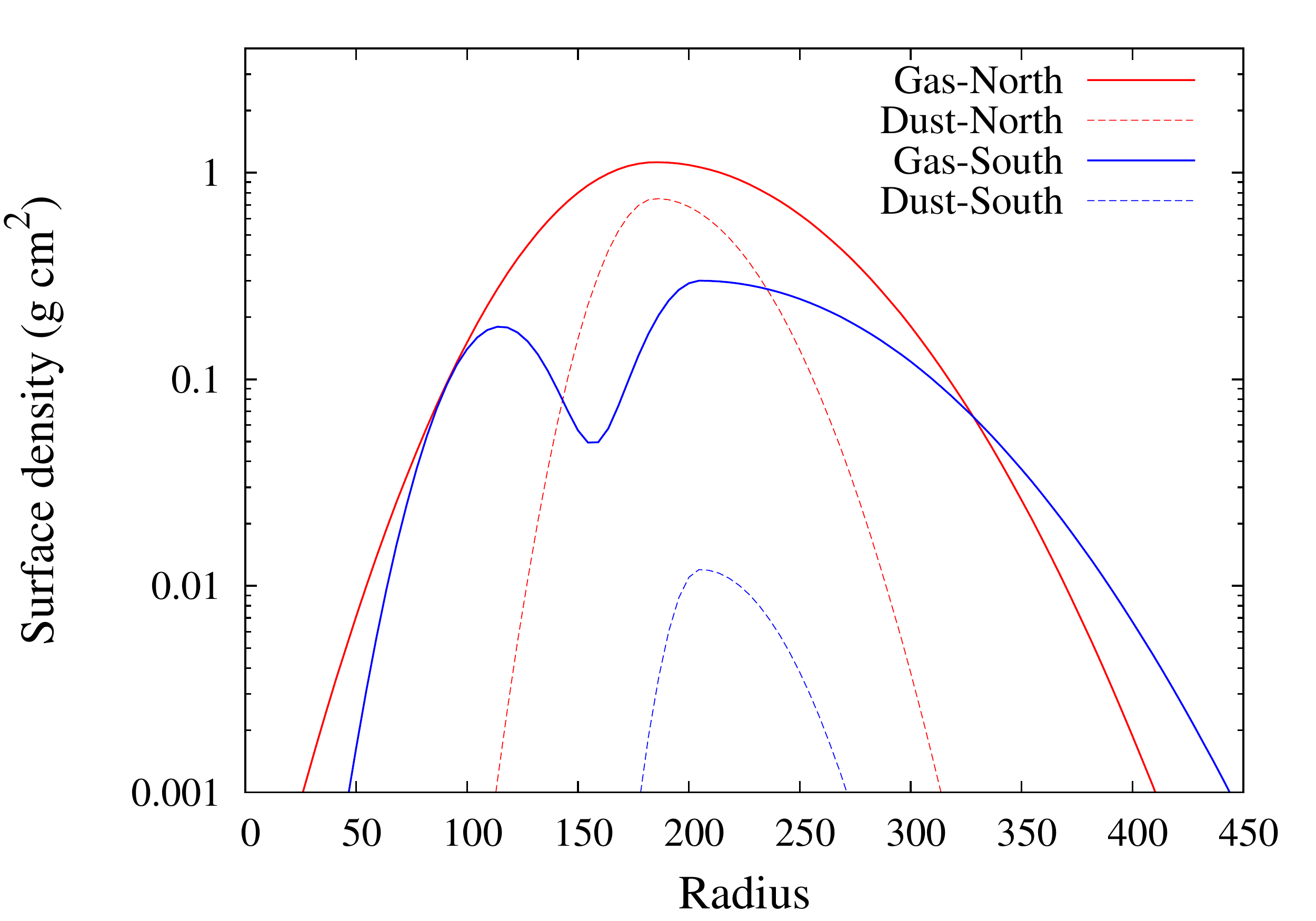}\\
  \caption{Gas and dust surface densities for the north and south profile.}
  \label{Fig:dens}
\end{figure}

The best fit model is shown in Figure~\ref{Fig:opticalD3}. The surface density $\Sigma_g(r)$ has a maximum at $r_r = 205$ AU, equal 
to that of the dust surface density, with values of $\Sigma_{g,0} = 0.30$ g cm$^{-2}$, $w_{g,in} = 30$ AU and $w_{g,out} = 100$ AU. 
The second Gaussian 
$\Sigma'_g(r)$ is centered at 115 AU, and is characterized by $\Sigma'_{g,0} = 0.18$ g cm$^{-2}$ and $w'_{g} = 30$ AU. 
This second Gaussian profile, while producing a visible CO emission, does not have a counterpart in the dust surface density. 
The summary of all the model parameters is shown in Table~\ref{tab:dens} and the surface densities in function of the radius 
in Figure~\ref{Fig:dens}.

\begin{table}  
  \begin{tabular}{|ccccc|}
    \hline
         & $\Sigma_{0}$ (g cm$^{-2}$) & $r_r$ (AU) &  $w_{in}$ (AU) & $w_{out}$ (AU) \\ \hline          
    \multicolumn{5}{l}{North (PA = 16\arcsec-26\arcsec)} \\ \hline
    Dust & 0.65  $\pm$ 0.05 & 185 $\pm$ 2 & 28  $\pm$ 2 & 50   $\pm$ 2 \\
    Gas  & 1.125 $\pm$ 0.1  & 185 $\pm$ 2 & 60  $\pm$ 5 & 85   $\pm$ 5 \\ \hline
    \multicolumn{5}{l}{South (PA = 216\arcsec-226\arcsec)} \\ \hline
    Dust & 0.012 $\pm$ 0.001 & 205 $\pm$ 2 & 17 $\pm$ 2 & 43  $\pm$ 2 \\
    Gas  & 0.18  $\pm$ 0.02 & 115  $\pm$ 5 & 30 $\pm$ 5 & 30  $\pm$ 5\\ 
         & 0.30  $\pm$ 0.02 & 205  $\pm$ 2 & 30 $\pm$ 5 & 100 $\pm$ 5\\ \hline    
  \end{tabular}
  \caption{Density prescription for the Gaussians describing the dust and gas surface density. On the south profile, 
         the gas surface density is described by 2 Gaussians.}     
  \label{tab:dens}
\end{table}

In conclusion, our analysis indicates that radial distribution of dust particles is well reproduced 
by asymmetric Gaussians whose inner widths are 2-3 times smaller than the outer widths. The 
density contrast between the dust crescent and the south side at the maximum of the dust surface 
density is $\sim$ 54. This is larger than the flux density contrast of 22 measured in the continuum and this is explained by 
the fact that the dust emission along the north profile is optically thick with a maximum optical depth $\tau_d$ = 1.9 
compared to an optical depth of 0.035 on the south. 
The gas distribution is more complex. The north profile is  well reproduced by a single asymmetric Gaussian centered 
at the same radius of the dust density. However, the distribution along the south is the sum of two Gaussian profiles 
separated by 90 AU.  Similarly to the dust, the gas surface density is the highest along the north direction. At the location 
of the maxima, the ratio between the north and south gas surface density is $\Sigma^{north}_{0,g}/\Sigma^{south}_{0,g}= 3.75$, 
while the ratio between the dust surface densities is $\Sigma^{north}_{0,d}/\Sigma^{south}_{0,d}=$ 54. 
Furthermore, the gas-to-dust ratio at $r=r_r$ is 1.7 on the north  and about 25 on the south. 
Similarly to the dust distribution, the inner widths for the gas distribution are smaller 
than the outer widths. Finally,  if we assume that the north and south profiles are representative of the two sides of the disk, 
the total dust mass is 1.5$\times$10$^{-3}$ M$_\odot$ and the total gas mass is 5.7$\times$10$^{-3}$ M$_\odot$. 
The disk mass is therefore about 7.2$\times$10$^{-3}$ M$_\odot$, or $\sim$ 0.3\% of the mass of the binary system. 
The measured gas-to-dust ratio is about 3-5 in the circumbinary disk, largely inferior to the standard ratio of 100.  

These results support the dust trapping scenario where dust is located at the radial and azimuthal gas pressure/density 
maximum. The differences between the inner and outer widths suggest, in the context of dust trapping, that the gas pressure gradient 
on the inner side is larger compared to the outer side. As discussed in the next section, the gas pressure maximum might originate from the 
interaction between the disk and the stellar companion, if in highly eccentric orbit, or a giant planet orbiting inside the inner cavity, 
 perhaps located 
at the position of source $c$ . Finally, the nature of the double Gaussian profile along the south profile is unclear.  We argue 
that the innermost Gaussian might be tracing the onset of a gas spiral. We note that this feature is located next to the 
shadow cast by the warped inner disk on the south side. Following \cite{Mont2016}, such a shadow may decrease the local gas
pressure and create trailing spirals. %HD~142527 is a particularly good case if such instabilities 
%must exist. The central Herbig star with a luminosity of 23 $L_{*}$ yields large contrast between the shadowed and the non-shadowed 
%region on the inner edge of the outer disk. 

% We tested, by reducing the dust scale height to a fifth of the gas scale height, if dust settling could explain it but the effect 
% was negligible. Another aspect that we neglected is that small grains of a few microns, not visible in our observations, must follow 
% the gas distribution, which extends to smaller radius. Indeed, more complex models could take into account different distributions 
% for small dust grains, coupled vertically and radially with the gas, and large grains, directly visible with ALMA observations in 
% the dust trap and settled into the disk. 

\section{Discussion}
\label{sec:discu}

\subsection{Origin of the central cavity}
\label{sec:cavity}

HD~142527 system possesses a very large cavity with a radius between 80 AU and 155 AU as measured from the dust 
thermal emission. However, $^{13}$CO J = 3-2 and C$^{18}$O J=3-2 lines emission is still visible in the millimeter-continuum 
cavity down to a radius of  $\sim$ 20 AU and 60 AU, respectively.  
Inside the dust cavity, we measure that dust surface density is at least 1200 lower compared to the value measured 
at the peak of the continuum emission (see section \ref{sec:Dust}). However, it is worth noting that ALMA observations 
mostly probe the distribution of millimeter grains. These grains have a Stokes number close to 1 and can be efficiently 
trapped in gas pressure bumps. On the contrary, much smaller grains, which only represent a small fraction 
of the dust mass, are expected to be coupled to the gas and therefore be accreted in larger amount down to a radius of at 
least 20 AU. This hypothesis is supported by the detection of near-infrared scattered light emission in the dust cavity down 
to an orbital radius of 70-80 AU \citep{Aven2014}. We argue that this trapping process is the origin of the difference between 
the spatial distribution of gas and dust inside the cavity and also of the low gas-to-dust ratio, of about 3-5, 
observed in the outer disk (see also \citealt{Zhu2012}). 

Given the binary nature of the HD~142527 system, it is plausible that the observed gas cavity is created by the gravitational 
interaction between the disk and the low-mass stellar companion. Recent simulations by \cite{Laco2016} have shown that the 
companion is probably orbiting on the inner disk plane, not coplanar with the outer disk, with an eccentricity of about 0.6 not 
sufficient to explain the dust cavity size. Furthermore, this interaction should lead to the formation of a gas density 
maximum very close to the cavity edge, i.e., between 40-60 AU. The measured location of the gas pressure maximum (185-205 AU) is 
about 4-5 times larger than predicted by the theory. In the case of eccentric orbits, both the radius of the cavity and the 
location of the gas density maximum are predicted to increase with the eccentricity remaining however quite close one to the other. 
The large ratio between the location of the gas density maximum and the cavity radius remains therefore an unsolved problem.

A way to reconcile theory and observations might be to invoke the presence of other, yet unseen, low mass companions orbiting 
between 13 AU and 180 AU. In principle, by carefully choosing the number, mass, and separation of these objects, it might be 
possible to obtain a gas surface density profile slowly increasing between 20-185 AU as that inferred for the north side of the 
disk \citep[Figure~\ref{Fig:dens}; similar multi-planet models are discussed in ][]{Isel2013,Dods2011}.  
One of such companion might be related to the source $c$, the compact source of continuum and CO emission detected 
at a distance of 50 AU from the central star (Figure~\ref{Fig:spec-strea}). The deviation of Keplerian rotation, inferred from 
the CO spectrum of source c, suggests that the emission might arise from material orbiting around a low mass companion. 
This is an intriguing possibility but observations at higher angular resolution and sensitivity are required to investigate the 
nature of this source and the presence of other compact objects in the HD~142527 system.  

Finally,  the disk photo-evaporation process might also be responsible for the opening of cavities in proto-planetary disks. 
However, according to \cite{Owen2011}, this process can not produce dust cavities larger than 20 AU with a relatively high 
accretion rate of about $10^{-7}$ M$_\odot$ yr$^{-1}$ as measured in HD~142527 \citep{Garc2006, Mend2014}. Also, this process 
equally affects dust and gas distributions but $^{13}$CO and C$^{18}$O are still observed inside the dust cavity. Our 
observations appear therefore to disfavor this scenario as the main cause for the observed disk morphology.

%According to \cite{Arty1994}, the cavity 
%created by a 0.25~M$_\odot$ companion should have a radius between 2-3 times its orbital radius, i.e., 30-60 AU assuming an 
%orbital radius between 15-20 AU, and might be eccentric. While the cavity radius and eccentricity predicted by these models is 
%consistent with our CO and dust continuum observations.

%(Lacour et al. 2016) separated the possible orbits for the companion in two families where one has its periapsis and
%apoapsis at PA coherent with our observations. Nevertheless, the presence of the stellar companion alone is
%surely not sufficient to explain the large size of the cavity and the distant radial position of the crest. Finally,
%no CO emission is detected at the position of the central compact continuum source.

%Another large system containing several close stars and a large cavity, difficult to explain only with the presence of the stars, is 
%GG Tau. \cite{Dutr2014} and \cite{Yang2017} found gas and dust streamers explaining the high accretion rate. 

% A simulation done by \cite{Casa2012} demonstrated that a 10 Jupiter-mass planet at an orbit of 90 AU could explain the cavity 
% shape. Nevertheless, such large planet would have been probably detected by IR observations \citep{Rame2012}. However, a smaller 
% planet of a few $M_j$ could also carve deep gaps in disks, especially if it has an eccentric orbit. 

\subsubsection{Origin of the lopsided disk}

HD~142527 has a very lopsided disk characterized by gas and dust surface density ratio of about 3.75 and 54, 
respectively, between the 
north and south side of the disk (Table~\ref{tab:dens}). The favored theory to produce such azimuthal asymmetries is the formation of a 
large vortex capable of concentrating dust grains. In a first step, small vortices can be triggered by the Rossby wave instability 
\citep{Love1999, Li2000} caused by radial variations in the gas density due to either a companion or, for example, a steep radial 
variation of the disk viscosity \citep{Rega2012}. Vortex formation might also result from hydrodynamic instabilities like the 
baroclinic instability \citep[see, e.g.,]{Owen2016}. Then, in a second step, they grow or merge together to form larger 
vortices \citep{Li2001}, which usually are limited to sizes of about one or two scale heights.

Different works have recently focused on scenarios able to generate a larger vortex with mode m = 1 surviving for a sufficiently long 
time (i. e. $\sim$ 1000 orbits) to produce an azimuthal dust concentration similar to those observed in HD 142527 and Oph IRS 48.
\cite{Mitt2015} proposed that a massive lopsided disk could displace the center of mass of the star+disk system creating an 
indirect force onto the gas particles that increases the vortex dimension and life expectancy \citep[see also][for a model where 
the indirect force effect is reduced by the disk self-gravity]{Zhu2016}. This hypothesis appears to be ruled out by our observations 
which show that the center of mass, calculated as the center of rotation of the disk, is located at the same position of the central 
compact dust continuum emission attributed to the circumprimary disk. In practice, our ALMA observations set an upper limit for the 
displacement between the primary star and the center of mass to about one tenth of the angular resolution of the observations, i.e., 
about 0.03\arcsec, or about 5 AU at the distance of this system. Furthermore, contrary to the model prediction, the central star is 
located closer to the dust crescent (165 AU) compared to the opposite side of the disk (180 AU). Therefore, the indirect force is not 
the driving factor of the disk asymmetry, probably due to the low disk mass, which is inferior to 1\% the mass of the central stars.

Large and long-lived vortices also require a low viscosity \citep{Fu2014}, which can not be easily reconciled 
with the relatively large mass accretion rate measured for HD~142527. A possible solution has been recently proposed by \cite{Zhu2016}. 
They studied the case of a large vortex created by the interaction between the disk and a 9 $M_{J}$ planet using a global MHD 
simulation. The vortex produces an azimuthal asymmetry of a factor of a few in the gas density and necessitates low viscosity 
corresponding to an equivalent Shakura-Sunyaev parameter $ \alpha <10^{-3}$. Such a low viscosity is obtained when ambipolar diffusion, 
the dominant non-ideal MHD term at large radii (i.e. 10s of AU), is added in the simulations as demonstrated in \cite{Zhu2014}. Also, 
a low viscosity help the triggering of RWI because the density perturbations induced by the planet are less smoothed out by diffusion. 
Following this model, ambipolar diffusion is less important in the innermost and more ionized regions of the disk. This leads to higher 
disk viscosity and consequently a larger mass accretion rate. Interestingly, \cite{Zhu2016} simulations  suggest that 
the disk-planet interaction might create an elliptical cavity ($e \sim 0.1-0.2$) more easily seen in the dust than in the gas 
emission. Moreover, the vortex is located closer to the star compared to the other side of the ring. 
This scenario is consistent with our observations and might explain the cavity shape without placing the companion or a planet 
on an eccentric orbit. 

Our analysis indicates that the gas-to-dust ratio at the center of the dust crescent is $\sim$ 1.7. This is very 
close to the minimum 
gas-to-dust ratio allowed by theoretical models which account for the dust feedback on the gas dynamics. Indeed, as the 
gas-to-dust ratio approaches unity, the dust particles, which tend to move at Keplerian velocity, will accelerate the  
sub-keplerian gas. As discussed in \cite{Fu2014}, this generates an instability that brakes the vortex in smaller dust clumps.  

Finally, we note that the concentration of dust particles toward an azimuthal gas pressure bump, e.g., a vortex, 
depends on the dust Stokes number $\tau_s$, which can be expressed as  \citep{Owen2016}
\begin{equation}
  \tau_s \approx \left ( \frac{a}{\mathrm{2~mm}} \right ) \left ( \frac{\mathrm{1~g~cm^{-2}}}{\Sigma_g} \right ) 
  \left ( \frac{\rho}{\mathrm{3~g~cm^{-3}}} \right ) ,
  \label{eq:opt}
\end{equation}
where $a$ is the size the dust grains and $\rho$ is the dust density. If we assume $\rho = 1.5$  
g~cm$^{-3}$ and take for the gas density the value at the center of the dust crescent (Table 1), $\Sigma_g = 1.125$ 
g~cm$^{-2}$, 
we obtain that the dust trapping is the most efficient ($\tau_s$ $=$ 1) for grains of $\sim$ 4.5 mm. Our observations 
probe the 
distribution of  dust particles with a size $a \sim \lambda/2\pi \sim 0.1$ mm, where $\lambda$ is the wavelength of the observation. 
This implies that observations at longer wavelengths, such as at 3 mm with ALMA, in addition to be more optically thin and better 
probe the disk density, should show a more compact concentration of dust particles. They might also tell if the dust crescent observed 
in the HD 142527 disk is a single coherent structure or consists instead of smaller clumps. An even higher dust concentration should 
be detected at cm wavelengths. So far, the only observations of HD142527 at wavelengths longer than 0.8~mm is a 9~mm continuum image 
obtained with the Australian Compact Array (ATCA)  \citep{Casa2015-2}. This image suggests indeed that millimeter size particles are 
more concentrated than sub-millimeter dust grains, but the low sensitivity of ATCA observations does not allow to study in 
details  the spatial distribution of the larger particles. 

%Australian Compact Array (ATCA). Such dust traps are a prime place to form planets. 
%Future dust density enhancement is possible via streaming instabilities which is especially efficient for large grains with 
%$\tau_s$ $\sim$ 1, able to concentrate dust pebbles on smaller scale which may eventually gravitationally collapse in planetesimals.

\section{Conclusion}
\label{sec:conclu}

We presented new ALMA observations of the HD~142527 system which resolve the 0.88~mm dust continuum, $^{13}$CO J=3-2, and C$^{18}$O 
J=3-2 line emission on a spatial scale of about 30-40 AU. By analyzing the morphology and kinematics of the gas and dust emission, and 
comparing the observations with theoretical disk models, we have reached the following conclusions:

\begin{enumerate}

  \item  HD~142725 circumbinary disk has a prominent horseshoe structure in the dust emission but much more azimuthally symmetric 
  when observed in CO emission lines. We have pointed out that this is partially due to the procedure usually adopted to image 
  molecular lines. In particular, we have demonstrated that continuum subtracted images of optically thick lines such as $^{13}$CO 
  J=3-2 may be misleading because a large fraction of the line emission is removed by the continuum subtraction if both line and 
  continuum are optically thick. This explains the observed anti-correlation between dust and gas in HD~142527, without requiring 
  any chemical or physical effect. We argue that continuum subtraction might also be the origin for the anticorrelation between dust and
  gas emission observed in  Oph IRS 48, SAO 206462, and SR~21.

  \item We have discovered a compact source of dust continuum and CO emission inside the dust depleted cavity at about
  50 AU from the primary star (source $c$). We argue that these emission might originate from material orbiting around a
  possible third object in this system. Future ALMA observations at higher angular resolution are required to better constrain
  the nature of this source.

  \item We compared the radial profile of the dust and CO emission to theoretical disk models to constrain the radial
  distribution of the circumstellar material. We found that along the direction intersecting the maximum of the dust
  emission (north direction), both dust and CO are well explained by Gaussian profiles centered at 185 AU from the primary 
  star. However, we find that the profile of the dust distribution is about a factor of 2 narrower than that of the gas.
  This result confirms that the observed horseshoe structure acts as a radial trap for millimeter size grains.
  Along the direction that intersect the minimum of the dust emission (south direction), we find that a double
  Gaussian profile is required to reproduce the observations. We argue that the innermost Gaussian might probe the
  innermost part of the spiral density wave observed at larger distances in the CO emission. 

  \item The comparison with theoretical models also indicates that the dust and CO densities vary by a factor of 54 and 
  3.75, respectively, between the minimum and maximum of the emission along the horseshoe structure. As a 
  consequence, we measure a gas-to-dust ratio of 1.7 at the center of the dust crescent. This is close to 
  the minimum gas to dust ratio allowed by theory before the
  back-reaction of dust grains affects the gas dynamics practically breaking the pressure bump apart. The large difference
  between the dust and gas azimuthal distribution indicates that the horseshoe structure traps dust particles azimuthally, as 
  well as radially. Our analysis also indicates that the mean value of the gas-to-dust ratio across this circumbinary disk is 
  between 3 and 5. This is much less than the typically assumed value of 100, suggesting that HD 142527 is more evolved 
  that usual protoplanetary disks. 

\end{enumerate}

Y.B. and A.I. acknowledge support from the NASA Origins of Solar Systems program through award NNX15AB06G.
A.I. acknowledges support from the NSF Grant No. AST-1535809. E.W. acknowledges support from the NRAO Student Observing Support Grant No. AST-
0836064.  J.M.C. acknowledges support from the National Aeronautics and Space Administration under Grant No. 15XRP15 20140 issued through 
the Exoplanets Research Program. This paper makes use of the following ALMA data  
\dataset[ADS/JAO.ALMA2012.1.00725.S]{https://almascience.nrao.edu/aq/?project_code=2012.1.00725.S}. ALMA is a partnership of 
ESO (representing its member states), NSF (USA) and NINS (Japan), together with NRC (Canada) and NSC and ASIAA (Taiwan) and KASI (Republic 
of Korea), in cooperation with the Republic of Chile. The Joint ALMA Observatory is operated by ESO, AUI/NRAO and NAOJ. The National 
Radio Astronomy  Observatory is a facility of the National Science Foundation operated under cooperative agreement by Associated 
Universities, Inc. This work has also made use of data from the European Space Agency (ESA) mission 
{\it Gaia}(\url{http://www.cosmos.esa.int/gaia}), 
processed by the {\it Gaia} Data Processing and Analysis Consortium (DPAC, \url{http://www.cosmos.esa.int/web/gaia/dpac/consortium}). 
Funding for the DPAC has been provided by national institutions, in particular the institutions participating in the {\it Gaia} 
Multilateral Agreement. The description of the Gaia mission is described in \cite{Gaiab2016} and the data release 1 in \cite{Gaiaa2016}.

\software{CASA (v4.3.10), RADMC-3D \citep{Dull2012}}
\facility{ALMA}

\appendix
\section{Dust scattering}
\label{app:A}

Radiative transfers and ray tracing performed on protoplanetary disks at millimeter wavelengths generally take into account the 
absorbing and emitting properties of the dust grains but neglect scattering. Recent observations of the circumbinary disk 
HD~142527 with ALMA by \cite{Kata2016} suggest that the detected polarized light might however come from thermal emission scattered 
by the dust grains themselves. Also, \cite{Kata2014} proposed that measuring the polarization degree in protoplanetary disks would 
bring additional constraints, along with the emission index $\beta$, on the dust grains size and porosity. However, the scattering 
opacity and phase function depend strongly on the chemical composition and internal structure of millimeter-size grains \citep{Taza2016} 
which are largely uncertain.  Consequently, it is difficult to properly investigate the effect of dust scattering on millimeter-wave 
continuum images.

% Furthermore, the theory to calculate optical opacities for composite grains is not fully understood (Bohren and Hoffmann 1983).
% of millimeter grains requires rigorous methods (T-Matrix, Rayleigh-Gans-Debye, etc) and the knowlegde of the hierarchical structure of the 
% dust aggregates to be precisely determined \citep{Taza2016}.}

\begin{figure*}[h]
  \includegraphics[angle=270,width=0.95\textwidth]{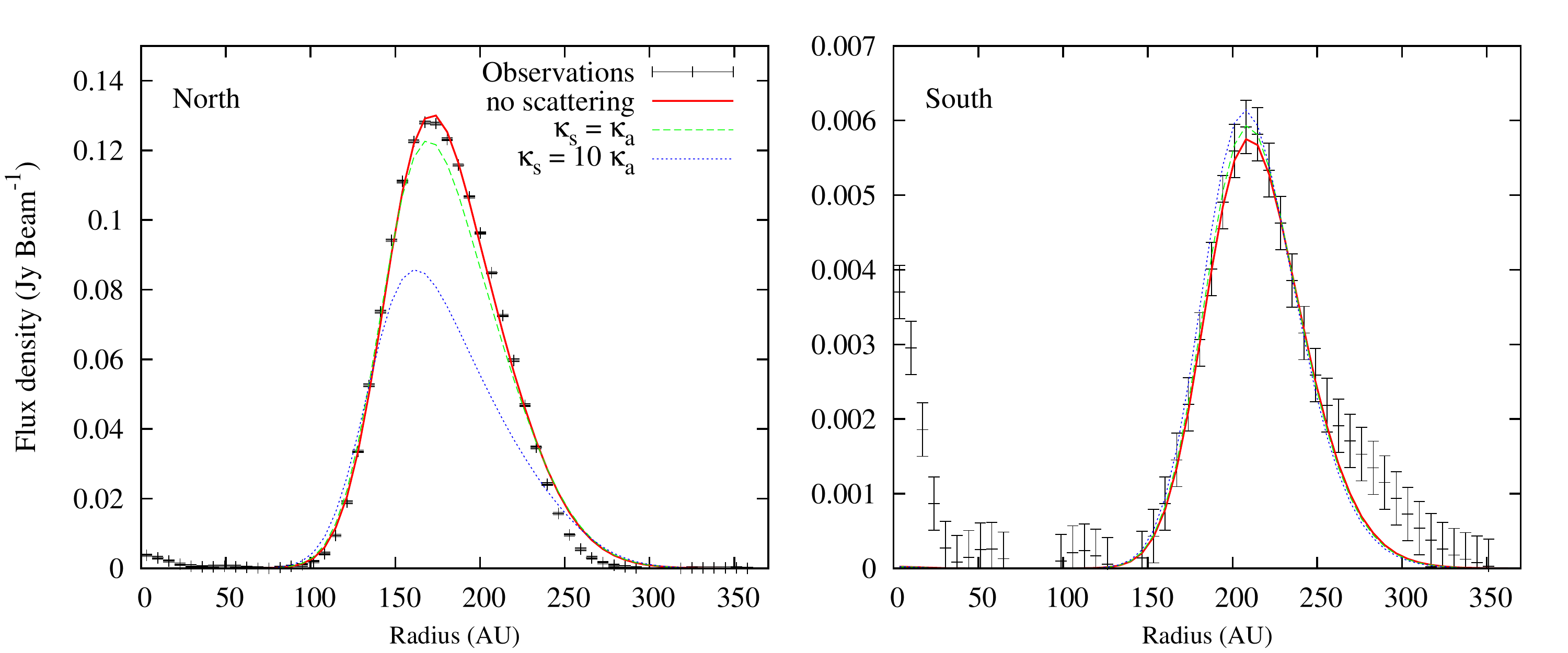}\\
  \caption{Radial cut along the north and south profiles at PA = 16$^{\circ}$-26$^{\circ}$ and 216$^{\circ}$-226$^{\circ}$, 
  respectively. The black crosses represent our observations. The red solid line is our prescription using no scattering, 
  the green dashed line corresponds to $\kappa_s$ = $\kappa_a$, and the blue dotted line to $\kappa_s$ = 10$\kappa_a$.}
  \label{Fig:sc}
\end{figure*}

As a simple approach, we show in Figure~\ref{Fig:sc} three models of the radial profile of the continuum emission along 
the north and south directions which assume an isotropic scattering phase function and a scattering opacity $\kappa_s$ equal to 
0, 1 and 10 times the absorption opacity $\kappa_a$ at the same wavelength. The dust radial distribution is given by our prescription which 
follows eq.~\ref{eq:dustpr} and the values in Table~\ref{tab:dens}. The models without scattering are indicated by the red solid lines and 
are equal to those shown in Figure~\ref{fig:opticalD1}. The blue dotted lines indicate the case $\kappa_s = 10~\kappa_a$ which is similar 
to that presented in M15. In this case, the scattering has a strong effect on the radial profile along the north direction, which
is optically thick, with a drop of 30\% in the maximum intensity. On the south profile, which is optically thin, the intensity only 
changes by 7\% or less. Our results are in agreement with those of \cite{Soon2017} who were not able to reproduce ALMA cycle 0 
data of HD~142527 with large scattering opacities due to the decrease of the maximum intensity, but instead used $\kappa_s$ $\sim$ $\kappa_a$. 
For this reason, we also represent the case $\kappa_s = \kappa_a$, in green dashed lines, and only find minor changes in flux within 5\%. 
This demonstrates that using moderate scattering opacities does not have important effects, 
at least with an isotropic phase function, in the determination of the dust surface density.

\end{document}